%% file: Adaptive.tex
\documentclass[%
]{mpi2015-cscpreprint}

\usepackage[american]{babel}

\usepackage{amssymb,amsfonts,amsmath}
\usepackage{subcaption}

\usepackage{float} 
\usepackage{amssymb,amsfonts,amsmath}
\usepackage{graphicx,color,changebar}
\usepackage{graphicx}    
\usepackage{bbm}
\usepackage{tikz}
\newlength\fheight
\newlength\fwidth
\usetikzlibrary{arrows,shapes,automata,backgrounds,petri,positioning}
\usetikzlibrary{decorations.shapes}
\usetikzlibrary{decorations.text}
\usetikzlibrary{decorations.fractals}
\usetikzlibrary{decorations.footprints}
\usetikzlibrary{shadows}
\usetikzlibrary{spy}
\usetikzlibrary{calc,patterns,
	decorations.pathmorphing,
	decorations.markings}

  \usepackage{algorithmic}
  \usepackage[linesnumbered,ruled,vlined]{algorithm2e}
  \SetKwInput{KwInput}{Input}                
  \SetKwInput{KwOutput}{Output}              
  
  \usepackage{cleveref}
  
  \newtheorem{Thm}{Theroem}
  \newtheorem{rem}{Remark}
  \newtheorem{defn}{Definition}[section]
  \newtheorem{prob}{Problem}[section]
  
  \usepackage{pgfplots}
  \newenvironment{customlegend}[1][]{%
  	\begingroup
  	\csname pgfplots@init@cleared@structures\endcsname
  	\pgfplotsset{#1}%
  }{  \csname pgfplots@createlegend\endcsname
  	\endgroup
  }
  
  \def\addlegendimage{\csname pgfplots@addlegendimage\endcsname}

\usepackage{mymacros}

\usetikzlibrary{calc,patterns,
	decorations.pathmorphing,
	decorations.markings}
\usepackage{todonotes}
\usepackage[normalem]{ulem}

\usepackage{amsmath}
\DeclareMathOperator*{\argmax}{arg\,max}
\renewcommand{\hat}[1]{\widehat{#1}}
\begin{document}
  


\title{A Greedy Data Collection Scheme For Linear Dynamical Systems}
  
\author[$1$]{Karim Cherifi}
\affil[$1$]{Institut f\"ur Mathematik MA 4-5, TU Berlin, Stra\ss e des 17. Juni 136, D-10623 Berlin, Germany\footnote{This work was done when the author was at the Max Planck Institute for Dynamics of Complex Technical Systems, Magdeburg, Germany.} \authorcr
	\email{cherifi@math.tu-berlin.de}, \orcid{0000-0003-1294-9291}}
	
\author[$2$]{Pawan Goyal}
\affil[$2$]{Max Planck Institute for Dynamics of Complex Technical Systems, Sandtorstra\ss e\ 1, 39106 Magdeburg, Germany\authorcr
	\email{goyalp@mpi-magdeburg.mpg.de}, \orcid{0000-0003-3072-7780}}
	
	\author[$3$]{Peter Benner}
	\affil [$3$]{Max Planck Institute for Dynamics of Complex Technical Systems, Sandtorstra\ss e 1, 39106 Magdeburg, Germany\authorcr
		\email{benner@mpi-magdeburg.mpg.de}, \orcid{0000-0003-3362-4103}}

\shorttitle{A greedy data collection scheme}
\shortauthor{K.Cherifi, P.Goyal, P.Benner}
  

\abstract{%
	Mathematical models are essential to analyze and understand the dynamics of complex systems. Recently, data-driven methodologies have got a lot of attention which is leveraged by advancements in sensor technology.
However, the quality of obtained data plays a vital role in learning a good and reliable model. Therefore, in this paper, we propose an efficient heuristic methodology to collect data both in the frequency domain and time-domain, aiming at the best possible information gain from limited experimental data. The efficiency of the proposed methodology is illustrated by means of several examples, and also, its robustness in the presence of noisy data is shown. 
	}
\maketitle

\section{Introduction} \label{sec:Intro}
In this paper, our focus is on inferring linear time-invariant (LTI) systems of the form
\begin{equation}\label{ss}
	\begin{aligned}
		\bE\dot \bx(t) &=\bA\bx(t) +\bB\bu(t), \\
		\by(t)&=\bC\bx(t)  + \bD \bu(t)
	\end{aligned}
\end{equation}
from data. Using the Laplace transform, the transfer function $\bH(s)$ of the system \eqref{ss} can be obtained  and is given as
\begin{equation}\label{TF}
	\bH(s)=\bC(s\bE-\bA)^{-1}\bB+\bD,
\end{equation}
where $s$ represents the frequency. The problem of learning dynamical systems of the form \eqref{ss} is widely studied and well understood, see, e.g., \cite{morMayA07,morDrmGB15,morDrmGB15a,PehW16,AntLI17,NakST18,yu2019non,brunton2019data}. 
A key ingredient to data-driven methodologies is data, and it is necessary to provide a good quality of it, ensuring a good quality of the inferred model. Typically, there are two possible ways to collect data, namely in frequency-domain and time-domain. In the frequency-domain, samples of transfer functions are given, whereas, in the time-domain, the output $\by$  is measured for an input $\bu$. Moreover, these data may be subject to noise. Depending on an experimental set-up, we may collect data in either of these domains. 

Among several widely used approaches, Loewner-based approaches \cite{morMayA07,AntLI17} to learn dynamical systems have gained increasing popularity. The fundamental idea of the Loewner-based approach is to construct a model that interpolates the given data points in the frequency domain. It has also been extended to time-domain data \cite{PehW16}. Furthermore,  the authors in \cite{AntLI17} have shown that if the minimum order of a system that realizes the system is $n$, then using $2n$ data points, one can construct the realization using the Loewner approach. However, the minimal order $n$ of a system is not known a priori; therefore, in practice, as many data as possible are collected, and the minimal order can be found using a truncation procedure, which is then followed by the construction of a minimal order realization. We refer the reader to ~\cite{morMayA07} for more details. 
However, this approach can impose practical constraints if collecting data is expensive. In this case, we aim to gather data points carefully and focus on collecting the data points to obtain as much information as possible about the underlying dynamics. In this direction, one may think of methods proposed in, e.g., \cite{morGugAB08,morPauA15,CheFB20,CheFRB20}, where interpolation points (or data) are chosen adaptively to obtain a good low-order or minimal realization of a high-fidelity model. However, these methods are not fully data-driven in the sense that they still require the high-fidelity model, which we here assume to be unknown.  Towards fully data-driven approaches, the method proposed in \cite{BeaGu12} determines the choice of measurement data in an optimal way (e.g., $\cH_2$-optimal).  But the method contains a few drawbacks -- these are: the method takes the order of the realization as an input, which is a hyper-parameter and needs to be chosen carefully. Secondly,  it may require measurements of the transfer function in the complex domain, but in an experimental set-up, we usually can only obtain measurements on the $\jmath\omega$-axis.

In this paper, we focus on inferring minimal linear dynamical systems using minimum possible measurement data, but without knowing the minimal order of the realization a priori and its realization in any form. For this, we discuss a heuristic approach, allowing us to select adaptively frequency points that provide the maximum information about the dynamics of the system. We also discuss the extension of the approach to time-domain data.

The paper is structured as follows. In \Cref{sec:Loewner}, the Loewner framework to learn linear dynamical systems is briefly presented. In \Cref{sec:Freq}, we present the heuristic approach to select frequency points (or transfer function measurements) to extract the maximum information about the system. Then, in \Cref{sec:Time}, we discuss an extension of the approach when input-output data are collected in the time-domain. In \Cref{sec:Numerics}, we illustrate the efficiency of the proposed greedy methodology and compare it with models that are obtained by taking measurements on a uniform grid. Finally, we conclude the paper with a summary and future research in \Cref{sec:conclusions}.

\section{Data-Driven Modeling Based on the Loewner Framework} \label{sec:Loewner}
 In this section, we briefly recap the Loewner approach \cite{morMayA07} to construct a realization from given transfer function measurements. To that end, for simplicity, we first write down the underlying realization problem for single-input single-output systems -- that is as follows:
\begin{prob} \label{prob:prob1}
	Given a set of interpolation points $Z := \{\sigma_1,\dots,\sigma_{2N}\}$ and corresponding transfer function measurements $\bH(\sigma_i)$, the goal is to identify a minimal realization whose rational transfer function is denoted by $\widehat\bH(s)$, satisfying the following interpolation conditions:
	\begin{equation}\label{IntCond}
		{\bH}(\sigma_i)=\widehat \bH(\sigma_i), \quad i \in \{1,\ldots, 2N\},
	\end{equation}	
	and the transfer function is given as:
	\begin{equation}\label{ApproxTF}
		\widehat{\bH}(s)=\widehat{\bC}(s\widehat{\bE}-\widehat{\bA})^{-1}\widehat{\bB}+\widehat{\bD},
	\end{equation}
	where $\hat\bE,\hat\bA \in \Rrr$, $\hat\bB \in \Rr$, $\hat\bC \in \Rr$ and $\hat\bD \in \R$ with $r$ being the order of a minimal realization. 
\end{prob}

This realization problem can be solved using the Loewner approach. For this, we need to split the measurement data into left and right measurements. Let us define the left and right interpolation points as $\lambda_j$ and $\mu_j$, where $j\in \{1,\ldots, N\}$, corresponding transfer function measurements are denoted by $\bH(\lambda_j)$ and $\bH(\mu_j)$, and $Z = \{\lambda_1,\ldots,\lambda_N\} \cup \{\mu_1,\ldots,\mu_N\}$. Having this data, in the following, we define the Loewner and shifted Loewner matrices.
\begin{defn}{\cite{morMayA07}} \label{def_loewnermatrices}
	Given left  data $\left(\lambda_j,\bH(\lambda_j)\right)$ and right data $\left(\mu_j,\bH(\mu_j)\right)$, the Loewner $\mathbb{L}$  and shifted Loewner $\mathbb{L}_{\mathrm s}$ matrices are defined as follows:
	\begin{align*}
		\mathbb{L} &=\left[ \begin{matrix}
			\frac{\bH(\lambda_1)-\bH(\mu_1)}{{{\lambda }_{1}}-{\mu_{1}}} & \cdots  & \frac{\bH(\lambda_n)-\bH(\mu_1)}{{{\lambda }_{n}}-{\mu_{1}}}  \\
			\vdots  & \ddots  & \vdots   \\
			\frac{\bH(\lambda_1)-\bH(\mu_n)}{{{\lambda }_{1}}-{\mu_{n}}} & \cdots  & \frac{\bH(\lambda_n)-\bH(\mu_n)}{{{\lambda }_{n}}-{\mu_{n}}}  \\
		\end{matrix} \right], \quad 
		\mathbb{L}_{\mathrm s}=\left[ \begin{matrix}
			\frac{{{\lambda }_{1}}\bH(\lambda_1)-{\mu_{1}}\bH(\mu_1)}{{{\lambda }_{1}}-{\mu_{1}}} & \cdots  & \frac{{{\lambda }_{n}}\bH(\lambda_n)-{\mu_{1}}\bH(\mu_1)}{{{\lambda }_{n}}-{\mu_{1}}}  \\
			\vdots  & \ddots  & \vdots   \\
			\frac{{{\lambda }_{1}}\bH(\lambda_1)-{\mu_{n}}\bH(\mu_n)}{{{\lambda }_{1}}-{\mu_{n}}} & \cdots  & \frac{{{\lambda }_{n}}\bH(\lambda_n)-{\mu_{n}}\bH(\mu_n)}{{{\lambda }_{n}}-{\mu_{n}}}  \\
		\end{matrix} \right].
	\end{align*}
\end{defn}

It was proven in \cite{morMayA07} that based on the Loewner and shifted Loewner matrices, a realization can be constructed that  solves  \Cref{prob:prob1}. This result is presented in the following theorem:
\begin{Thm}(\cite{morMayA07}) \label{Thm:Thm1}
	Given left data $(\lambda_j,\bH(\lambda_j))$ and right  data $(\mu_j,\bH(\mu_j))$, consider the Loewner and shifted Loewner matrices as defined in \Cref{def_loewnermatrices}.  Moreover, assume that the feed-through term $\bD$ is known. Then, an interpolating realization can be constructed as follows:
	\begin{equation}\label{eq:real}
		\begin{split}
			\widehat{\bE} &= \mathbb L, \quad \widehat{\bA} = \mathbb L_s - \mathbbm 1\bD\mathbbm 1^\top, \quad
			\widehat{\bB} = \bV - \mathbbm 1\bD, \\ \widehat{\bC}& = \bW - \bD\mathbbm 1^\top, \quad \widehat{\bD} = \bD. 
		\end{split} 
	\end{equation}
	where $\mathbbm 1 \in \R^N$ is the vector of ones. Furthermore, the realization \eqref{eq:real}  is minimal, assuming $\mathbb L$ is of full-rank. 
\end{Thm}
Furthermore, in case the Loewner matrix $\mathbb L$ is singular, there exists a lower order realization of order $\hn < n$, where $\hn = \rank{\mathbb L}$, satisfying the interpolation conditions. To remove the redundant information and obtain the minimal realization, a compression step based on the SVD of $[\mathbb L,\mathbb{L}_{\mathrm s}]$ can be employed, see, e.g., \cite{morMayA07,AntLI17}. For simplicity, we have presented the realization theory using the Loewner approach for SISO systems; however, it can be extended to multi-input multi-output using tangential interpolation. We refer to \cite{morMayA07, AntLI17} for more details. We also remark that the realization \eqref{eq:real} can be complex if data are complex, but if the interpolation points are closed under conjugation, then there exists an orthogonal transformation that allows to determine the real realization using orthogonal transformation. 

The success of the Loewner approach has been shown in various applications, see, e.g., \cite{IonAnt14,morIonA14,PouKer20}. Its success lies in the quality of data and assumes that there are enough data for a wide range of frequencies. However, suppose the data collection process is expensive. In this case, it is essential to make a smart choice to measurements to obtain maximum information about the system with limited data. With this aim, in the following section, we discuss an adaptive scheme to collect measurements to extract maximum information about the system. 

\section{A Greedy-based Data Collection Scheme} \label{sec:Freq}
Naturally, the quality of collected data is a key ingredient to learning a good and reliable model, describing the underlying dynamics of a process. Obviously, one can collect as much data as possible if feasible. It can then be followed by inferring a system realization by using, e.g., the Loewner approach \cite{morMayA07}. The approach involves a compression step that squeezes the information lying in all the given data. 
However, a mass collection could be expensive in many scenarios. Therefore, in this section, we discuss a greedy scheme that can guide us to collect the data so that we can expect to extract as much new information about the dynamical system from every measurement as possible. In the following, we first note down a corresponding problem. 
\begin{prob}\label{prob:problem1}
	Identify the underlying state-space dynamical system of a process using minimal number of transfer function evaluations.
\end{prob}
\Cref{prob:problem1} has been discussed in the literature in different ways. Towards this, it has been shown in \cite{AntLI17} that the minimum number of transfer function measurements needed to determine a minimal state-space realization of order $n$ is  $2n$. And such a realization can be determined by employing the Loewner approach \cite{morMayA07}. However, if only data is given, we cannot know a priori the minimal order of state-space realization that captures the dynamics of the process. Therefore, typically, one aims at collecting $N \gg 2n$ data points, which is then followed by a compression step to determine the minimal order of the state-space model that captures the dynamics of the process. 
Furthermore, a step towards collecting good data, the authors in \cite{BeaGu12}, presented an approach inspired by the IRKA algorithm \cite{morGugAB08} that focused on using interpolation points that facilitate obtaining a model, minimizing the $\cH_2$ error between the learned model and the unknown ground-truth model. However, again, the approach needs the order of the state-space realization as input, which may not be known in advance. Furthermore, we mention that there also exist other approaches that aim at determining or collecting measurements greedily but require a ground-truth high-fidelity model, see, e.g.,  \cite{BedBDR20,morPauA15,CheFB20,CheFRB20}.

On the other hand, our goal is to construct a minimal realization of a process using as a minimum number of measurement data as possible, where we neither know the order of a minimal realization in advance, nor have access to any high-fidelity model. 
Furthermore, one can theoretically consider taking the transfer function measurements on the complex domain, and often, it is known that good, in fact optimal, measurement points are complex as shown in \cite{morGugAB08,BeaGu12}. However, from an experimental viewpoint and physical interpretation of a transfer function, we assume that transfer function measurements are taken anywhere on the imaginary $\jmath\omega$-axis. The motivation of considering measurements on the $\jmath\omega$-axis are:
\begin{itemize}
	\item[(a)] A transfer function at $\jmath\omega$ can be estimated by exciting the system using an input containing sine and cosine functions of the frequency $\omega$. It can be extended to multiple frequencies as well. 
	\item[(b)] From the control theory perspective, a transfer function can be better interpreted on the $\jmath\omega$-axis, for example, using the Bode plot. More details can be found in, e.g., \cite{Ogata10}.
\end{itemize}

To that end, we first assume that we have an initial rough construct of a realization that still can be far from satisfactory. An initial rough model can be constructed using only a few measurement points.  We denote the transfer function of the initial realization by $\bH_{\text{init}}(s)$. Furthermore, let us denote the transfer function of the unknown ground truth realization by $\bH_{\text{true}}(s)$.
In this case, we ideally would like to update the model using the  point on the $j\omega$-axis, where the maximum error between the true and initial realized systems occurs; hence, we select the interpolation point that solves the following optimization problem:
\begin{equation} \label{eq:opt_prob}
	\sigma_{\texttt{new}} := \underset{\sigma \in \jmath\omega}{\mathop{\arg\max }}\,\| \bH_{\texttt{true}}(\sigma)-{\bH_\texttt{init}}(\sigma)\|,
\end{equation}
where $\omega \in \R. $ However, to solve the optimization problem \eqref{eq:opt_prob}, we require the transfer function of the ground-truth realization, which is not available. Therefore, we seek for an alternative approach, allowing us to estimate good interpolation points in an iterative procedure.  
For this, let us assume that we have $2k$ measurement points, to begin with, and denote these measurements by tuples $(\sigma_i,\bH(\sigma_i))$, $i \in \{1,\ldots,2k\}$, where $\sigma_i$ and $\bH(\sigma_i)$ are the interpolation points and the transfer function measurements at these points. Before we proceed further, we define the following notation:
\begin{itemize}
	\item $\Sigma_i := \{\sigma_1,\ldots,\sigma_{2i}\}$,
	\item  The transfer function of the realization constructed using interpolation points $\mathrm \Sigma_{m}$ is denoted by $\bH_m(s)$, and the order of the constructed realization is $m$. 
\end{itemize}
Next, we construct realizations using interpolation points $\Sigma_{k-1} \subset \Sigma_k$ and  $\Sigma_k$, which are respectively denoted by $\bH_{k-1}(s)$ and $\bH_k(s)$. As discussed earlier, we ideally would add a new measurement at the frequency that solves the problem \eqref{eq:opt_prob}.
%
Since we do not know $\bH_{\text{true}}$, in the following, we discuss a heuristic approach to relax the  optimization problem. First note that
\begin{align}
	\|\bH_{\text{true}}(\sigma)-{\bH_{k}}(\sigma)\|  &= \|\bH_{\text{true}}(\sigma)- {\bH_{k-1}}(\sigma) -\left( {\bH_{k}}(\sigma) - {\bH_{k-1}}(\sigma) \right)\| \nonumber\\ 
	&\leq \|\bH_{\text{true}}(\sigma)- {\bH_{k-1}}(\sigma)\| + \| {\bH_{k}}(\sigma) - {\bH_{k-1}}(\sigma) \|. \label{eq:k_eqn}
\end{align}
With unknown $\bH_{\text{true}}(\sigma)$, we rather focus on defining our next measurements data based on the second part of  \eqref{eq:k_eqn} with a constraint;  that is -- we should exclude the regime of already taken measurements points.  Hence, we solve an optimization problem to obtain a frequency point at which transfer function measurement needs to be taken to obtain more information about the underlying dynamics as follows:
\begin{equation}\label{eq:relaxed_opt}
	\max_{\sigma \in \jmath \omega} \bg(\sigma) \|{\bH_{k}}(\sigma) - {\bH_{k-1}}(\sigma) \|,
\end{equation}
where the function $\bg(\sigma)$ can be thought of as a mask that aims at excluding the regime of points at which measurements have already been collected. If $\bg(\sigma)$ is known, then we can solve \eqref{eq:relaxed_opt} to obtain our new measurement points which possibly bring the most information about the system as they would occur at the maximum change in the transfer function from the previous to the next step.

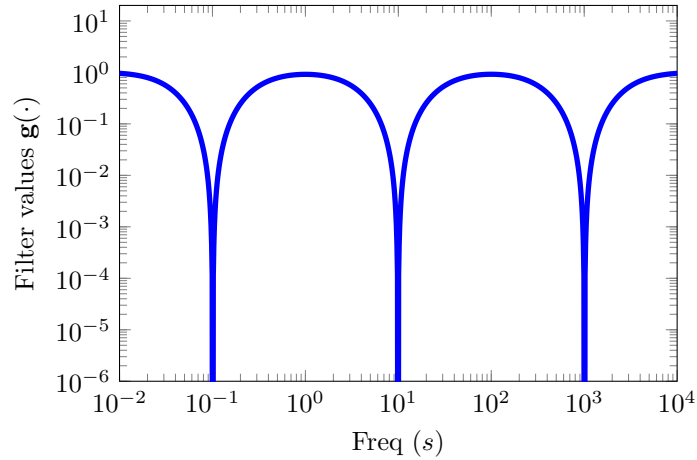
\begin{figure}[tb]
	\centering
	\setlength\fheight{5.0cm}
	\setlength\fwidth{0.5\textwidth}
	\input{Figures/testFilter.tikz}
	\caption{A visual illustration of the filter when the measurements are already taken at the frequencies $\sigma_i=\{10^{-1}, 10^{1}, 10^{3}\}$.}
	\label{fig:Filt}
\end{figure}

Naturally, the choice of the function $\bg(\cdot)$ plays an important role in determining next measurement points. So, in the following, we discuss a choice of the mask function $\bg(\sigma)$. As discussed, a choice of the mask function should be such that it excludes the regime of already considered data points. Among many possible choices, here, we propose the following choice:
\begin{equation*}
	\bg(\sigma) = \prod _{i=1}^{2k}\tilde \bg(\sigma,\sigma_i),
\end{equation*}
where $ \tilde\bg(\sigma,\sigma_i)$ is defined as: 
\begin{equation}  \label{eq:Filt}
	\tilde\bg(\sigma,\sigma_i)= 1-{\mathrm e}^{(-\beta(\log(|\sigma| + \epsilon)-\log(|\sigma_i| + \epsilon))^2)},
\end{equation}
with  $\sigma_i, i \in \{1,\ldots,2k\}$, being already considered measurement points, and $\epsilon$ and $\beta$  are positive constants and hyper-parameters.  It can be noticed that $\bg(\sigma)$ is zero at all the considered points with small values in their neighborhood in order to favor the exploration over the already chosen interpolation points. The mask function can also be seen as a notch filter at multiple frequencies and $\beta$ can be seen as the bandwidth of the filter. To illustrate the mask function, we plot the function for $\beta=0.6$, $\epsilon=10^{-15}$ and the notch frequencies  $\sigma_i=\{10^{-1}, 10^{1}, 10^{3}\}$ in \Cref{fig:Filt}.
It clearly shows that the mask takes smaller values near the considered frequency points. 

Next, we discuss how to further simplify the choice of the function $\bg(\sigma)$. First, note that $\bH_k(\sigma_i) = \bH_{k-1}(\sigma_i)$ for $i\in \{1,\ldots,{2k-2}\}$ assuming that the realization is determined by the Loewner approach, and thus, the interpolation at the given measurement points is exact. As a result, we can also consider the optimization problem to determine our next measurement point as follows:
\begin{equation} \label{eq:Sn}
	\sigma_{2k+1} := \argmax_{\sigma \in \jmath \omega} \prod_{i = 2k-1}^{2k}\tilde\bg(\sigma,\sigma_{i})\| {\bH_{k}}(\sigma) - {\bH_{k-1}}(\sigma) \|.
\end{equation} 

In this paper, we utilize the Loewner approach to construct the realization. Therefore,  it is preferred to have an even number of measurement points to avoid the construction of a rectangular realization. Hence,  it is desired to include two new additional points at each step. Consequently, in order to choose one additional point, we solve the following optimization problem once $\sigma_{2k+1}$ is obtained:
\begin{equation} \label{eq:Sn2}
	\sigma_{2k+2} := \argmax_{\sigma \in \jmath \omega} \prod_{i = 2k-1}^{2k+1}\tilde\bg(\sigma,\sigma_{i})\| {\bH_{k}}(\sigma) - {\bH_{k-1}}(\sigma) \|.
\end{equation}
As a result, we have two new frequency points $\sigma_{2k+1}$ and $\sigma_{2k+2}$ at which we take transfer function measurements. Thus, we update our realization $\bH_{k+1}$ by using measurements at $\sigma_i$, $i \in \{1,\ldots,2k+2\}$. In the SISO case, the norm between the transfer function measurements just becomes the absolute value.

We repeat the procedure until: 
\begin{equation} \label{eq:converge}
	\max\limits_{s\in j\omega}\|\bH_k(s) - \bH_{k-1}(s)\| \leq \texttt{tol}.
\end{equation}

We sketch the whole procedure in Algorithm \ref{alg:Alg1}. When the condition \eqref{eq:converge} is met, the algorithm converges as we have obtained the minimal realization of the system and no new information can be added to model. 

The algorithm is not sensitive to small changes in the input hyperparameters. For example, small changes of $\beta$ in the mask function will not have a big impact on the convergence of the algorithm as long as the mask function keeps the properties described above. This is shown in the numerical section. The other hyperparameter \texttt{tol} needs to be set by the user as a trade off between accuracy of the resulting model and the number of interpolation points used.

\begin{algorithm}[!tb]
	\DontPrintSemicolon
	
	\KwInput{A parameter $\beta$, and initial measurement points $\Sigma_k := \{\sigma_1,\ldots, \sigma_{2k}\}$.}

	Determine realizations $\bH_{k-1}$ and $\bH_k$ by using measurement points $\Sigma_{k-1} \subset \Sigma_k$ and $\Sigma_k$, respectively and by employing the Loewner approach. 

Compute $\texttt{Err}:= \max\limits_{s\in j\omega}\|\bH_k(s) - \bH_{k-1}(s)\| $.

	\While{	\texttt{$\texttt{Err}> \texttt{tol}$} }
	{	
		Compute the new measurement point $\sigma_{2k+1}$ according to the optimization problem \eqref{eq:Sn}.
		
		Deduce $\sigma_{2k+2}$ according to \eqref{eq:Sn2} .
		
		Estimate the transfer function at the measurement points $\sigma_{2k+1}$ and $\sigma_{2k+2}$. 
	
		Determine a realization using measurement points at $\Sigma_k \cup \{\sigma_{2k+1},\sigma_{2k+2}\}$.
		
		$k \leftarrow k+1$.
	
		Compute $\texttt{Err} := \max\limits_{s\in j\omega}\|\bH_k(s) - \bH_{k-1}(s)\| $.
}
	\KwOutput{A learned model, whose transfer function is given by $\bH_{k}$. }
	\caption{A greedy selection of measurement points in the frequency-domain.}
	\label{alg:Alg1}
\end{algorithm}

\section{Extension to Time Domain Data} \label{sec:Time}
In the previous section, we have discussed a greedy approach for frequency domain measurements. Here, we suppose that we have access to the time-domain data instead of the frequency-domain data. In this case, we need to design an input to collect time domain data so that we extract the maximum information about the process. In this case, we can also employ Algorithm~\ref{alg:Alg1} with a slight modification. First, note that the authors in \cite{PehGW17} have proposed a methodology to realize a state-space model using time-domain data, where frequency-domain measurements are estimated using time-domain data. This is followed by obtaining a realization using the Loewner framework described in \Cref{sec:Loewner}. However, the methodology heavily depends on the choice of input, and the choice should be made in such a way that it allows us to describe the dynamics of the system completely. Hence, in this section, we discuss a suitable choice of inputs. Using an extension of Algorithm \ref{alg:Alg1}, one can adaptively choose frequency points composing an input.
In practice, time-domain measurements are typically collected at a regular interval; hence, the measurements are discrete. Therefore, the proposed greedy procedure needs to be adapted in a discrete setting. In what follows, we discuss the adaptation that needs to be made to allow us to design an input to extract as much information as possible.  Note that the interpolation points are no longer on the $s-$plane but are rather on the unit circle of the $z-$plane. Thus, the frequency range on the $\jmath\omega$-axis needs to be mapped on the unit circle using an appropriate discretization method.
In order to obtain time-domain measurements, the system is excited with an input spanning a set of interpolation points and the system response is measured. We design an input $\bu_p^{(k)}$ using a sum of sine and cosine functions, i.e., the input in the $k${th} step of the algorithm at time $\bT_sp$, where $\bT_s$ is a sampling time and $p$ is a non-negative integer, can be given as:
\begin{equation} \label{eq:input}
	{\bu_p^{(k)}}=\frac{1}{K}\sum\limits_{l=1}^{2}{(1+\jmath)}\left(\cos \left(\sigma_{2k+l}\right)+\jmath \sin\left(\sigma_{2k+l}\right)\right),
\end{equation}
where $\jmath := \sqrt{-1}$, $p \in \{0,\ldots,K-1\}$, $K$ is the number of time-domain measurements. 

Following the same strategy as in \Cref{sec:Freq}, the goal is to estimate frequency-domain data, construct a realization using the interpolation points and corresponding transfer function estimates. However, since only time domain data are available, one has first to use this data to estimate the frequency measurements.

The complete procedure is as follow. Given an initial realization $\bH_{\text{init}}(s)$, the next two interpolation points  $\sigma_{2k+1}$ and $\sigma_{2k+2}$ are computed as in \eqref{eq:Sn} and \eqref{eq:Sn2}. Then, we construct the input $\bu_p^{(k)}$ (at step $k$ of the algorithm and time step $p$) as in \eqref{eq:input} with the two interpolation points $\sigma_{2k+1}$ and $\sigma_{2k+2}$. We simulate the system to obtain an output having a value of ${\by_p^{(k)}}$ at the step $k$ of the algorithm and at time step $p$. Assembling this time-domain data, one can then compute estimates for the transfer function  $\bH_{\texttt{true}}(\sigma_{2k+1})$ and $\bH_{\texttt{true}}(\sigma_{2k+2})$. These values are computed using a least-squares problem \cite{PehGW17} as follows:
\begin{equation}
	\label{LSprob2}
	\underset{\widehat{\bH}\in {{\mathbb{C}}^{r}}}{\mathop{\arg \min }}\,\left\| \bF\widehat{\bH}-\bar{\by} \right\|_{2}^{2},
\end{equation} 	
where $\bF\in {{\mathbb{C}}^{(\bK-{{k}_{\min }})\times 2}}$ and $\widehat{\bH}$ are as follows :

\begin{equation}\label{F}
	\bF=\left[ \begin{matrix}
		{{\bU}_{1}}{e}^{\jmath\sigma_{2k+1}{{{k}_{\min }}}} &  {{\bU}_{2}}{e}^{\jmath\sigma_{2k+2}{{{k}_{\min }}}}  \\
		\vdots  & \vdots   \\
		{{\bU}_{1}}{e}^{\jmath\sigma_{2k+1}{(K-1)}} & {{\bU}_{2}}{e}^{\jmath\sigma_{2k+2}{(K-1)}},
	\end{matrix} \right], \quad\text{and}  \quad
	\widehat{\bH}=\left[	\begin{matrix}
		\widetilde{\bH}(\sigma_{2k+1}) \\
		\widetilde{\bH}(\sigma_{2k+2})
	\end{matrix} \right],
\end{equation}
where $\widetilde{\bH}(\sigma)$ is an estimate of $\bH_{\texttt{true}}(\sigma)$, and the output vector is defined as
$\bar{\by}:={{\left[{\by_{{k}_{\min }}^{(k)}},\ldots,{\by_{K-1}^{(k)}}\right]}^{\top}}$ 
such that $\by_p^{(k)}$ is the measurement of the output at time step $p$ and the step $k$ of the algorithm. $\bU_1$ and $\bU_2$ are the nonzero (discrete) Fourier transform components of the input $\bu$, given in \eqref{eq:input} which will be, in fact, at frequencies $\sigma_{2k+1}$ and $\sigma_{2k+2}$, respectively. Moreover, $k_{\min}$ is chosen such that the system reaches a steady state (approximately) after $k_{\min}$ time steps. For a proof of the derivation of the least-squares problem \eqref{LSprob2}, we refer to the discussion in \cite{PehGW17}.

The next steps are then similar to the steps of Algorithm \ref{alg:Alg1}. A realization is constructed using all the estimated points at $\Sigma_k \cup \{\sigma_{2k+1},\sigma_{2k+2}\}$. The new interpolation points are then computed as in \eqref{eq:Sn} and \eqref{eq:Sn2}. The whole procedure is repeated until the error between two iterations is small enough, i.e., \[\max\limits_{\|z\|=1}\|\bH_k(z) - \bH_{k-1}(z)\| \leq \texttt{tol}.\]


\begin{rem}
	Although the realization is constructed using all collected interpolation points, the input is constructed at each step using only two interpolation points. It results in a considerable reduction in the computation time of the least-squares problem \eqref{LSprob2} compared to the case where all interpolation points have to be inferred at once.
\end{rem}

\section{Numerical experiments} \label{sec:Numerics}
In this section, we illustrate the proposed methodology by means of three examples. The first two examples, namely, the penzl \cite{Penzl06} and the beam \cite{morChaV02} examples, consider frequency-domain measurements.  In the last example, time-domain data is used to construct a realization for an RLC circuit \cite{morGugA03} with an adaptive choice of measurements. We compare our proposed approach, where we make a careful choice of measurement points, with the Loewner approach, where logarithmically equidistant measurement points are considered. In both cases we choose the best splitting method of right and left interpolation points as discussed in \cite{morKarGA19a}. 
In the following, we also note some details. Typically, we are interested in a frequency range $[\omega_{min},\omega_{max}]$, and we consider $M$ points, on a logarithmic scale, from that given range, denoted by $\bQ := \left[q_1,\ldots,q_{M}\right]$. Whenever a measurement point is taken, we assume that it is taken for the set $\bQ$. 
Moreover, the Loewner framework \cite{morMayA07} is utilized to construct a realization using measurement points and transfer function evaluations. 
The initial realization is constructed using $6$ approximately logarithmically equidistant interpolation points taken from the set $\bQ$. This is done by computing first $6$ logarithmically equidistant interpolation points  $\left[\widehat q_1,\ldots,\widehat q_{6}\right]$ in the frequency range  $[\omega_{min},\omega_{max}]$ and then find the interpolation points $\left[\tilde q_1,\ldots,\tilde q_{6}\right]$ that are used in the algorithm by solving the following optimization problem:
\begin{equation}
	\underset{\tilde q_{i} \in \bQ}{\mathop{\arg\min }}\,\|\tilde q_{i}-\widehat q_i\|.
\end{equation}
%
In each step of the algorithm, the interpolation points are organized into an interlacing manner []. The filter parameters in \eqref{eq:Filt} are set to $\beta=0.6$ and $\epsilon=10^{-15}$. The tolerance in Algorithm~\ref{alg:Alg1} is set to $\texttt{tol}=10^{-8}$.
All the numerical experiments are done on an AMD Ryzen 7 PRO 4750U processor~CPU@1.7GHz, up to 8MB cache, 16 GB RAM, Ubuntu 20.04 LTS, MATLAB Version 9.8.0.1323502(R2020a) 64-bit(glnxa64).

\subsection{Penzl example}
\begin{figure}[tb]
	\centering
	\begin{tikzpicture}
		\begin{customlegend}[legend columns=3, legend style={/tikz/every even column/.append style={column sep=0.5cm}} ,legend cell align={left}, legend entries={Ground-truth, Equidistant Loewner,  Adaptive interpolation points, Adaptive Loewner, Equidistant interpolation points}, ]
			\addlegendimage{color=blue, line width = 2pt}
			\addlegendimage{color=magenta,dashdotted, line width = 2pt}
			\addlegendimage{color=black, only marks, mark=x, mark options={solid, black}, line width = 1.5pt}
			\addlegendimage{color=green!50!black,dashed, line width = 2pt}
			\addlegendimage{color=red, only marks, mark=asterisk, mark options={solid, red}, line width = 2pt}
		\end{customlegend}
	\end{tikzpicture}
	\setlength\fheight{5.0cm}
	\setlength\fwidth{0.8\textwidth}
	\input{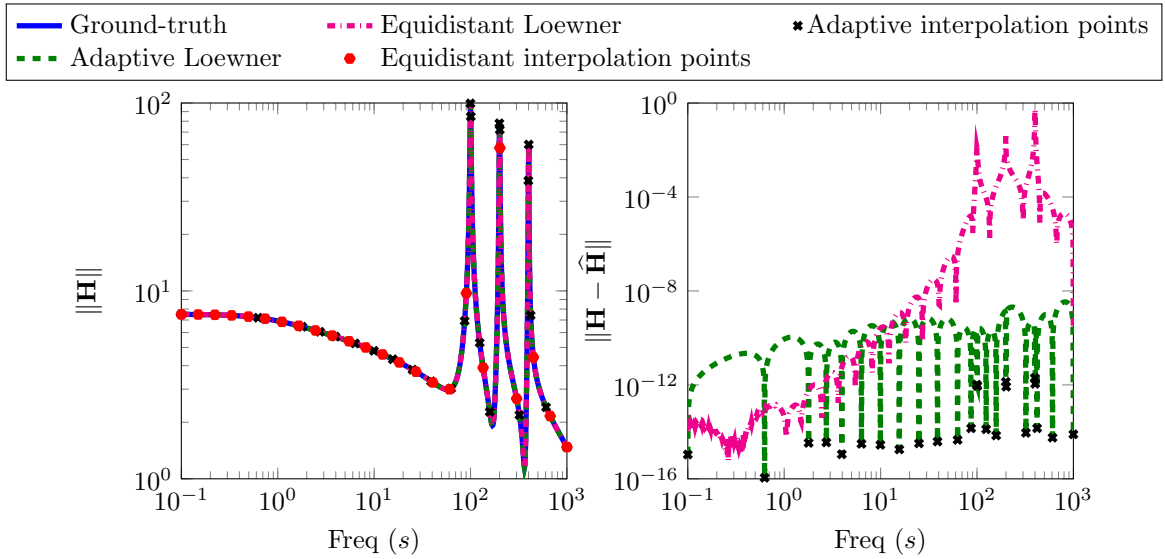}
	\caption{Penzl example: The Bode plot of the ground-truth, adaptively generated system and a realization with equidistant points and the corresponding error between them.}
	\label{fig:FomBode}
\end{figure}

As a first example, we consider the penzl example \cite{Penzl06} of order $N=1~006$, which is referred to as the FOM example in the \texttt{SLICOT} model reduction benchmark \cite{morChaV02}. We are interested in the frequency range $[10^{-1},10^{3}]$. Using  Algorithm \ref{alg:Alg1}, we obtain a realization that on termination uses $24$ interpolation points.  For comparison, we also construct a realization using $24$ logarithmically equidistant interpolation points in the given frequency range. We compare these two learned realizations in \Cref{fig:FomBode}, where we observe that our greedy scheme to collect measurements clearly outperforms the scheme when measurements are taken equidistantly. Algorithm \ref{alg:Alg1} focuses on the region where the transfer function is involved, and it automatically trends to add more points around the peaks of the transfer function. On the other hand, the realization, constructed using the logarithmically equidistant points, has a larger error as it fails to capture these peaks accurately.

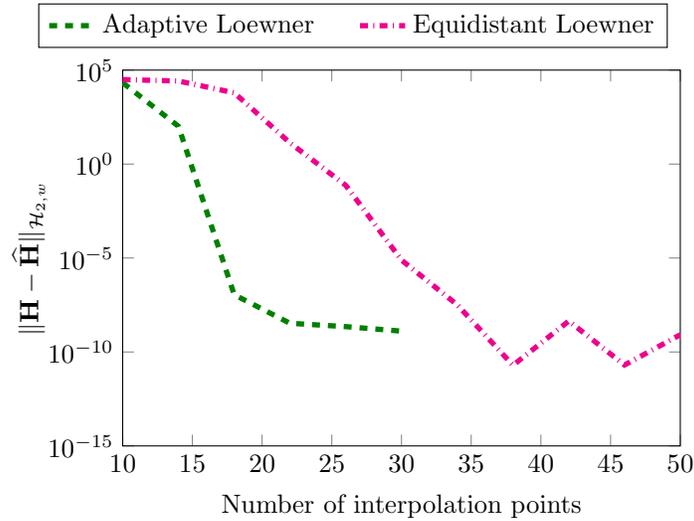
\begin{figure}[tb]
	\centering
	\begin{tikzpicture}
		\begin{customlegend}[legend columns=3, legend style={/tikz/every even column/.append style={column sep=0.5cm}} , ,legend cell align={left}, legend entries={Adaptive Loewner, Equidistant Loewner}, ]
			\addlegendimage{color=green!50!black,dashed, line width = 2pt}
			\addlegendimage{color=magenta,dashdotted, line width = 2pt}
		\end{customlegend}
	\end{tikzpicture}
	\setlength\fheight{5.0cm}
	\setlength\fwidth{0.5\textwidth}
	\input{Figures/Exm_fom_H2norm.tikz}
	\caption{Penzl example: A comparison between the frequency-limited $\cH_2$-norm error for the adaptively chosen points and the equidistant points. }
	\label{fig:FomH2}
\end{figure}
\begin{figure}[tb]
	\centering
	\begin{tikzpicture}
		\begin{customlegend}[legend columns=2, legend style={/tikz/every even column/.append style={column sep=0.3cm}} , ,legend cell align={left}, legend entries={ Adaptive Loewner $\beta=0.1$, Adaptive Loewner $\beta=0.6$, Adaptive Loewner $\beta=3$, Equidistant Loewner}, ]
			\addlegendimage{color=blue, line width = 2pt}
			\addlegendimage{color=green!50!black,dashed, line width = 2pt}
			\addlegendimage{color=magenta,dashdotted, line width = 2pt}
			\addlegendimage{color=red, only marks, mark=asterisk, mark options={solid, red}, line width = 0.8pt}
		\end{customlegend}
	\end{tikzpicture}
	\setlength\fheight{5.0cm}
	\setlength\fwidth{0.4\textwidth}
	\input{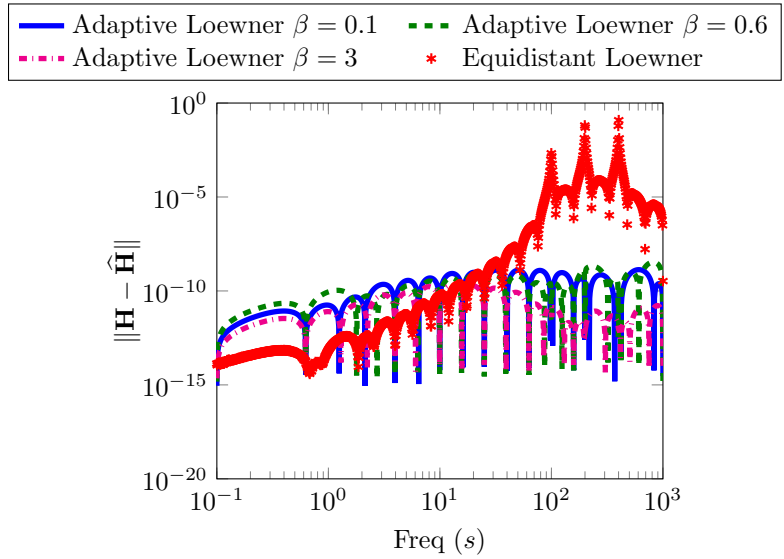}
	\caption{Penzl example: A comparison between the resulting bode plot for different values of $\beta$ used to defined the filter $\bg(\cdot)$ in \eqref{eq:Filt}. }
	\label{fig:Fombw}
\end{figure}

Furthermore, in \Cref{fig:FomH2}, we compare the frequency-limited $\cH_2$-norm \cite{morWil70} of the learned and ground-truth systems in terms of the number of interpolation points used.
%
%
%
We use the \texttt{MORLAB} software package for the computation of the frequency-limited norm~\cite{morBenW20}. The figure shows that the frequency-limited $\cH_2$ error decays faster when the measurements are collected adaptively (Algorithm \ref{alg:Alg1}).
Finally, we study the robustness of Algorithm 1 with respect to the hyper-parameter $\beta$. We report the quality of the learned realization with respect to the parameter $\beta$ in  \Cref{fig:Fombw}. We note that Algorithm \ref{alg:Alg1} is quite robust to the parameter, and all learned realizations outperform the one obtained using the logarithmically equidistant measurement points.  

\subsection*{Measurement noise}
Many studies have been made about the robustness of the Loewner framework to noise, see, e.g., \cite{morLefIA10,morDrmP19,KerFCD18}. We do a preliminary study and observe the performance of our approach under noisy measurements.  We conduct three experiments, where we corrupt the transfer function measurements with Gaussian white noise of standard deviation $\sigma=[10^{-4},10^{-5},10^{-6}]$, respectively. We compare our approach with a realization based on logarithmically equidistant interpolation points and measurements corrupted with the same noise level as for the adaptive case. The $\cH_2$-norms of learned and the ground-truth models are compared in \Cref{Table:H2fom}. The table shows that even though the $\cH_2$ error for the adaptive algorithm increases with the level of noise, it yields much better models as compared to the one obtained using the logarithmically equidistant points. This shows that the adaptive method is superior even in the presence of noise in the measurements. It is important to note that the Loewner framework was used to illustrate the application of our method. In the noisy data case, there maybe better methods where our method can be used.

\begin{table}[tb!]
	\centering
	\begin{tabular}{| c | c | c | c | c |}
		\hline
		& \multicolumn{4}{c}{ Standard deviation of noise ($\sigma$) } \vline \\ \hline
		Method & 0 & $10^{-6}$ & $10^{-5}$ & $10^{-4}$ \\ \hline
		Adaptive  &	$3.287\cdot10^{-9}$ & $1.544\cdot10^{-7}$ & $1.788\cdot10^{-5}$ & $5.886\cdot10^{-4}$ \\ \hline
		Equidistant  & $7.737\cdot10^{-6}$ & $2.400\cdot10^{-3}$ & $1.400\cdot10^{-2}$ & $3.259\cdot10^{-1}$ \\ \hline	
	\end{tabular}
	\caption{Penzl example: A comparison of the frequency-limited $\cH_2$-norm of the error between the ground-truth and realized systems under different levels of noise in the measurement data.}
	\label{Table:H2fom}
\end{table}

\subsection{Beam example}
\begin{figure}[tb!]
	\centering
	\begin{tikzpicture}
		\begin{customlegend}[legend columns=3, ,legend cell align={left}, legend style={/tikz/every even column/.append style={column sep=0.25cm}} , legend entries={Ground-truth, Equidistant Loewner,  Adaptive interpolation points,  Adaptive Loewner, Equidistant interpolation points}, ]
			\addlegendimage{color=blue, line width = 2pt}
			\addlegendimage{color=magenta,dashdotted, line width = 2pt}
			\addlegendimage{color=black, only marks, mark=x, mark options={solid, black}, line width = 1.5pt}
			\addlegendimage{color=green!50!black,dashed, line width = 2pt}
			\addlegendimage{color=red, only marks, mark=asterisk, mark options={solid, red}, line width = 2pt}
		\end{customlegend}
	\end{tikzpicture}
	\setlength\fheight{5.0cm}
	\setlength\fwidth{0.9\textwidth}
	\input{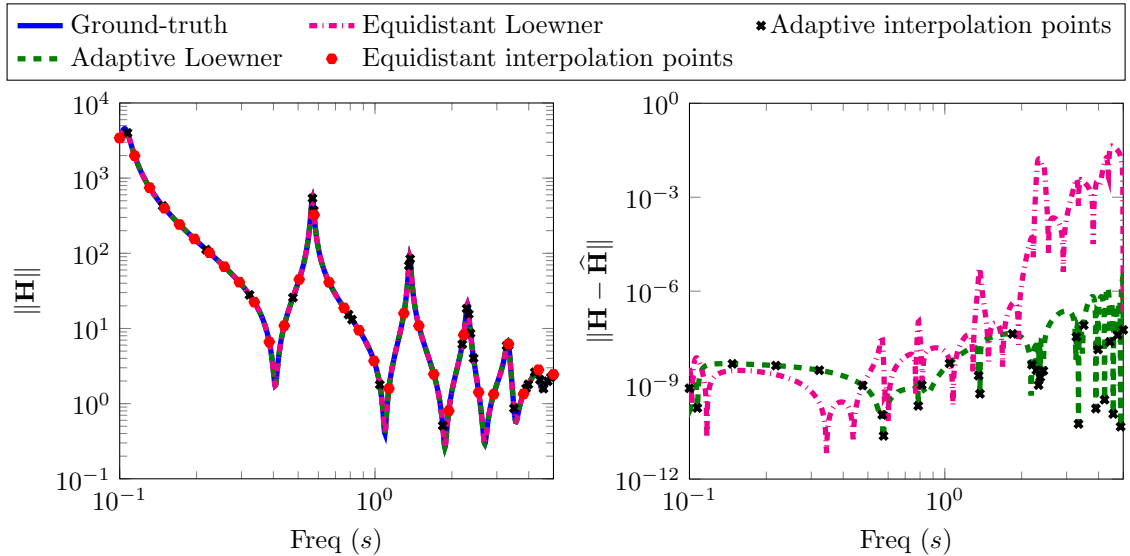}
	\caption{Beam Example: The Bode plot of the ground-truth system, adaptively generated system, and a realized system with equidistant points. The right figure shows the corresponding error between the ground-truth and realized systems.}
	\label{fig:BeamBode}
\end{figure}

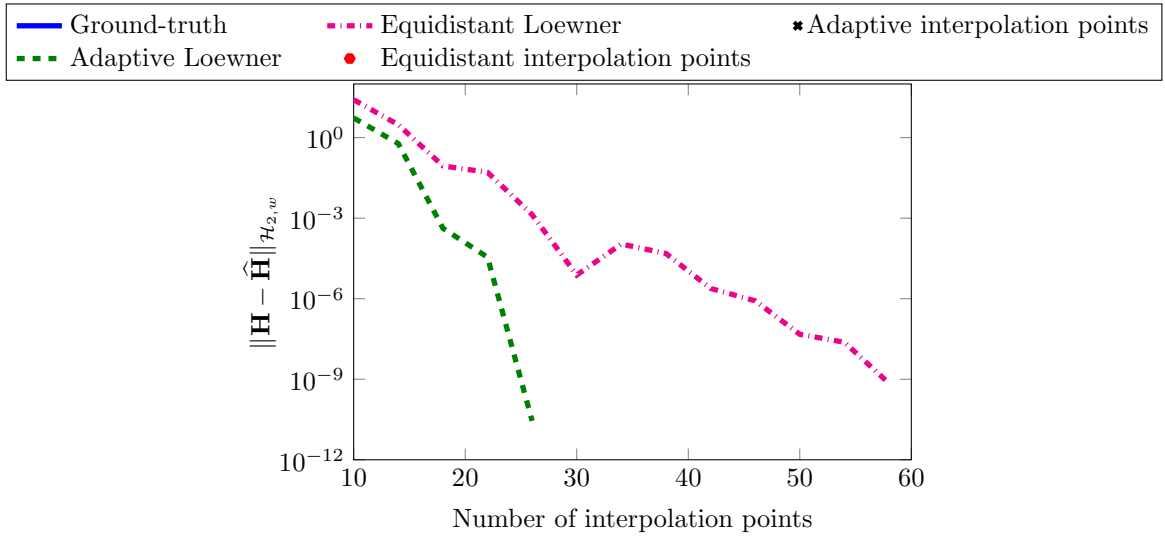
\begin{figure}[tb!]
	\centering
	\begin{tikzpicture}
		\begin{customlegend}[legend columns=3, legend style={/tikz/every even column/.append style={column sep=0.5cm}} ,legend cell align={left}, legend entries={Ground-truth, Equidistant Loewner,  Adaptive interpolation points, Adaptive Loewner, Equidistant interpolation points}, ]
			\addlegendimage{color=blue, line width = 2pt}
			\addlegendimage{color=magenta,dashdotted, line width = 2pt}
			\addlegendimage{color=black, only marks, mark=x, mark options={solid, black}, line width = 1.5pt}
			\addlegendimage{color=green!50!black,dashed, line width = 2pt}
			\addlegendimage{color=red, only marks, mark=asterisk, mark options={solid, red}, line width = 2pt}
		\end{customlegend}
	\end{tikzpicture}
	\setlength\fheight{5.0cm}
	\setlength\fwidth{0.5\textwidth}
	\input{Figures/Exm_beam_H2norm.tikz}
	\caption{Beam Example: A comparison between the frequency-limited $\cH_2$ norm error for the adaptively chosen points and the equidistant points. }
	\label{fig:BeamH2}
\end{figure}

In this example, we consider the beam model from the SLICOT model reduction library \cite{morChaV02}. It comes from the discretization of a PDE \cite{morAntSG01}. The input is the force applied at one end, and the output is the displacement resulting from the applied force at another end. We consider the frequency range of operation $[10^{-1},5]$. Algorithm \ref{alg:Alg1} is applied, and a  realization is obtained using 30 interpolation points as shown in \Cref{fig:BeamBode}. As in the first example, we also construct a realization using 30 logarithmically equidistant logarithmic scale interpolation points distributed over the chosen range of frequencies. Here again, we notice that most of the interpolation points using the adaptive method are chosen around the areas where most changes are happening. In contrast, measurements taken at logarithmically equidistant points yield a realization having a larger error. In \Cref{fig:BeamH2}, we observe that adaptively chosen interpolation points decrease the $\cH_2$-norm error faster as compared to the equidistant ones. 

\subsection{RLC circuit}
\begin{figure}[tb!]
	\centering
	\begin{tikzpicture}
		\begin{customlegend}[legend columns=3, legend style={/tikz/every even column/.append style={column sep=0.5cm}} ,legend cell align={left}, legend entries={Ground-truth, Equidistant Loewner,  Adaptive , Adaptive Loewner}, ]
			\addlegendimage{color=blue, line width = 2pt}
			\addlegendimage{color=magenta,dashdotted, line width = 2pt}
			\addlegendimage{color=green!50!black,dashed, line width = 2pt}
		\end{customlegend}
	\end{tikzpicture}
	\setlength\fheight{5.0cm}
	\setlength\fwidth{0.9\textwidth}
	\input{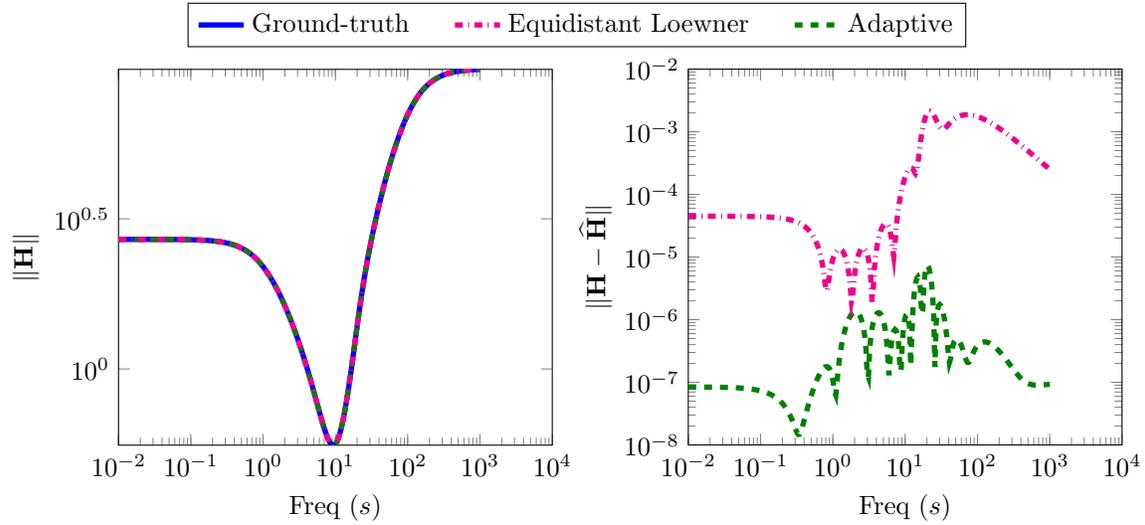}
	\caption{RLC example: A comparison for the Bode plot of the ground-truth and the identified models. }
	\label{fig:RLCBode}
\end{figure}

\begin{figure}[tb!]
	\centering
	\begin{tikzpicture}
		\begin{customlegend}[legend columns=3, legend style={/tikz/every even column/.append style={column sep=0.5cm}} ,legend cell align={left}, legend entries={Ground-truth, Equidistant Loewner,  Adaptive , Adaptive Loewner}, ]
			\addlegendimage{color=blue, line width = 2pt}
			\addlegendimage{color=magenta,dashdotted, line width = 2pt}
			\addlegendimage{color=green!50!black,dashed, line width = 2pt}
		\end{customlegend}
	\end{tikzpicture}
	\setlength\fheight{5.0cm}
	\setlength\fwidth{0.6\textwidth}
	\input{Figures/Exm_RLCSerkan_time_step.tikz}
	\caption{RLC example: A comparison between the output response of the ground-truth and the identified models. }
	\label{fig:RLCstep}
\end{figure}
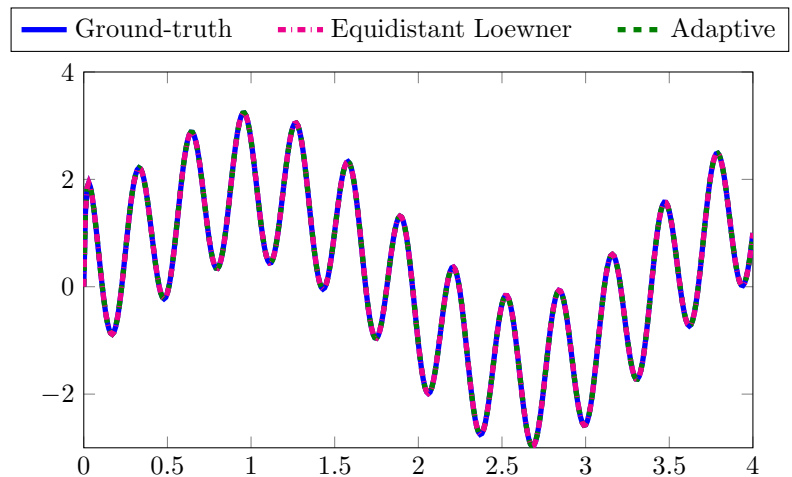

Finally, we consider an RLC circuit with 100 resistors, capacitors and inductors \cite{morGugA03,BGV20}. The frequencies are taken within the range of frequencies $[10^{-2},10^{3}]$. In order to obtain input/output time-domain measurements, a system simulation using the input \eqref{eq:input} is performed to obtain the output in each step. Applying the procedure described in \Cref{sec:Time} directly realizes a model of order $12$. As in the first example, we consider the realization of a model using $12$ logarithmically equidistant interpolation points and their complex conjugates on the unit circle in order to ensure that the resulting system is real, as discussed in \cite{PehGW17}. A comparison of the Bode plots is shown in \Cref{fig:RLCBode} and a comparison between the output response of the ground-truth system and its realization is shown in \Cref{fig:RLCstep} using the input \[\displaystyle{ \bu(t)=\sin(2t)+\sin(20t)}.\]
Our algorithm recovers the model with a relatively low error, and the model captures the time-domain response of the ground-truth model. It shows that our method provides a promising direction for the long-standing subject of input choice for time-domain system identification.

\section{Conclusion}\label{sec:conclusions}
In this paper, we have presented a purely data-driven realization method that greedily chooses measurement points to extract maximum information about the dynamics of a system. We have discussed both cases where data are taken in frequency and time-domain. We have illustrated the efficiency of the proposed methodology by means of three benchmark examples. It was shown that our method outperforms the case where the measurement data are taken at logarithmically equidistant points in a given frequency range. 

\bibliographystyle{unsrt}
\bibliography{mor}
  
\end{document}

%% file: Figures/testFilter.tikz
%
%
\definecolor{mycolor1}{rgb}{0.00000,0.44700,0.74100}%
\begin{tikzpicture}

\begin{axis}[%
width=0.951\fwidth,
height=\fheight,
at={(0\fwidth,0\fheight)},
scale only axis,
xmode=log,
xmin=0.01,
xmax=10000,
xminorticks=true,
ymode=log,
ymin=1e-6,
ymax=20,
yminorticks=true,
xlabel = {Freq $(s)$},
ylabel = {Filter values $\bg(\cdot)$},
axis background/.style={fill=white}
]
\addplot [color=blue, line width = 2pt]
  table[row sep=crcr]{%
0.01	0.9592489426752\\
0.0101392540755882	0.957661634354322\\
0.0102804473209331	0.956022592253322\\
0.010423606739764	0.954330580403261\\
0.010568759711848	0.952584352301413\\
0.0107159339982267	0.950782651671977\\
0.0108651577465254	0.948924213267395\\
0.0110164594963366	0.947007763710692\\
0.0111698681846782	0.945032022379228\\
0.0113254131515281	0.942995702330178\\
0.0114831241454351	0.940897511268008\\
0.0116430313292088	0.938736152554156\\
0.011805165285688	0.936510326259078\\
0.0119695570235904	0.934218730256716\\
0.0121362379834424	0.931860061361438\\
0.0123052400435926	0.929433016507374\\
0.0124765955263087	0.926936293970036\\
0.012650337203959	0.924368594630027\\
0.0128264983052806	0.921728623278569\\
0.0130051125217341	0.919015089964482\\
0.0131862140139475	0.916226711382223\\
0.0133698374182495	0.913362212300424\\
0.0135560178532937	0.910420327030381\\
0.0137447909267754	0.907399800933771\\
0.0139361927422414	0.904299391968863\\
0.0141302599059953	0.90111787227434\\
0.0143270295340983	0.897854029789788\\
0.0145265392594678	0.894506669911833\\
0.014728827239075	0.891074617184755\\
0.0149339321612425	0.887556717024387\\
0.0151418932530435	0.88395183747396\\
0.0153527502878042	0.880258870990471\\
0.0155665435927106	0.876476736260085\\
0.0157833140565212	0.872604380040939\\
0.016003103137387	0.868640779031645\\
0.0162259528707809	0.864584941763714\\
0.0164519058775366	0.860435910515971\\
0.0166810053720006	0.856192763249\\
0.0169132951702965	0.851854615557507\\
0.0171488196987054	0.847420622638449\\
0.0173876240021625	0.842889981272623\\
0.0176297537528721	0.838261931817393\\
0.0178752552590424	0.833535760208064\\
0.0181241754737424	0.828710799965396\\
0.0183765620038817	0.823786434206613\\
0.0186324631193156	0.818762097657214\\
0.0188919277620767	0.813637278660791\\
0.0191550055557353	0.808411521183997\\
0.0194217468148903	0.803084426813719\\
0.0196922025547917	0.797655656743455\\
0.0199664245010979	0.792124933745812\\
0.020244465099768	0.786492044127999\\
0.0205263775270925	0.780756839667096\\
0.0208122156998634	0.774919239521871\\
0.021102034285686	0.768979232117819\\
0.0213958887134342	0.762936877002093\\
0.0216938351838518	0.756792306664919\\
0.0219959306803007	0.750545728324097\\
0.0223022329796594	0.744197425669099\\
0.0226128006633728	0.737747760561318\\
0.0229276931286565	0.731197174686949\\
0.0232469705998565	0.724546191159002\\
0.0235706941399673	0.717795416064933\\
0.0238989256623105	0.71094553995637\\
0.024231727942376	0.70399733927743\\
0.0245691646298279	0.696951677728138\\
0.0249113002606779	0.689809507559455\\
0.0252582002696278	0.682571870796495\\
0.0256099310025846	0.675239900386484\\
0.0259665597293487	0.667814821268122\\
0.0263281546564802	0.660297951359002\\
0.0266947849403432	0.652690702457827\\
0.0270665207003324	0.644994581058218\\
0.0274434330322837	0.637211189070991\\
0.0278255940220712	0.629342224451829\\
0.0282130767593947	0.62138948173139\\
0.0286059553517574	0.613354852444987\\
0.0290043049386399	0.605240325459028\\
0.0294082017058706	0.597047987191589\\
0.0298177229001967	0.588780021724545\\
0.0302329468440578	0.580438710804814\\
0.0306539529505653	0.572026433732448\\
0.0310808217386906	0.563545667133374\\
0.0315136348486648	0.554998984614771\\
0.0319524750575921	0.546389056301212\\
0.032397426295282	0.537718648249854\\
0.0328485736603004	0.528990621743117\\
0.0333060034362459	0.520207932457462\\
0.0337698031082509	0.511373629507066\\
0.0342400613797143	0.502490854361355\\
0.0347168681892656	0.493562839635538\\
0.0352003147279668	0.484592907753504\\
0.0356904934567523	0.475584469482603\\
0.0361874981241128	0.466541022340054\\
0.0366914237840249	0.457466148870921\\
0.0372023668141307	0.448363514797799\\
0.03772042493417	0.439236867042566\\
0.03824569722467	0.430090031620786\\
0.0387782841458946	0.420926911409531\\
0.0393182875570577	0.411751483789631\\
0.0398658107358044	0.402567798163588\\
0.0404209583979631	0.393379973350583\\
0.0409838367175726	0.384192194860259\\
0.0415545533471888	0.375008712047153\\
0.0421332174384729	0.365833835147905\\
0.0427199396630678	0.356671932203573\\
0.043314832233764	0.347527425869598\\
0.0439180089259609	0.338404790116216\\
0.0445295850994266	0.329308546822292\\
0.045149677720361	0.320243262265782\\
0.0457784053837662	0.311213543514269\\
0.0464158883361278	0.302224034719159\\
0.0470622484984128	0.293279413317415\\
0.0477176094893875	0.284384386144838\\
0.0483820966492596	0.275543685465127\\
0.0490558370636505	0.266762064919162\\
0.0497389595879006	0.258044295399082\\
0.0504315948717136	0.249395160851982\\
0.0511338753841433	0.240819454018159\\
0.0518459354389291	0.232321972109043\\
0.0525679112201842	0.223907512430097\\
0.0532999408084409	0.215580867954108\\
0.0540421642070592	0.20734682285045\\
0.0547947233690029	0.199210147976022\\
0.0555577622239888	0.191175596333691\\
0.0563314267060135	0.183247898504175\\
0.0571158647812643	0.175431758057437\\
0.0579112264764176	0.167731846949704\\
0.0587176639073326	0.160152800912355\\
0.0595353313081437	0.15269921483898\\
0.0603643850607587	0.14537563817696\\
0.061204983724767	0.138186570329992\\
0.062057288067765	0.131136456078006\\
0.0629214610961034	0.124229681020968\\
0.0637976680860628	0.117470567053047\\
0.0646860766154633	0.110863367873668\\
0.0655868565957143	0.104412264541946\\
0.0665001803043112	0.0981213610809628\\
0.0674262224177834	0.0919946801383509\\
0.0683651600451024	0.0860361587095769\\
0.0693171727615541	0.080249643930272\\
0.0702824426430835	0.0746388889438941\\
0.0712611543011175	0.0692075488509144\\
0.0722534949178721	0.0639591767456437\\
0.0732596542821523	0.0588972198466957\\
0.0742798248256492	0.0540250157269842\\
0.0753142016597437	0.0493457886490102\\
0.0763629826128224	0.044862646011069\\
0.0774263682681127	0.0405785749098487\\
0.0785045620020451	0.0364964388247405\\
0.0795977700231499	0.0326189744289989\\
0.080706201411495	0.0289487885327219\\
0.0818300681586739	0.0254883551624211\\
0.0829695852083491	0.022240012781754\\
0.0841249704973612	0.0192059616577829\\
0.0852964449974103	0.0163882613769024\\
0.0864842327573172	0.0137888285143534\\
0.0876885609458743	0.0114094344610042\\
0.0889096598952916	0.00925170341083996\\
0.0901477631452492	0.0073171105123484\\
0.0914031074875623	0.00560698018673317\\
0.0926759330114688	0.00412248461562977\\
0.0939664831495469	0.00286464240072265\\
0.0952750047242729	0.0018343173973971\\
0.0966017479952265	0.00103221772427516\\
0.097946966706954	0.000458894950210702\\
0.099310918137498	0.000114743460027165\\
0.100693863147603	9.99996786810367e-16\\
0.102096066230605	0.000114743403794459\\
0.103517795563018	0.000458894499276374\\
0.104959323055823	0.00103221619632509\\
0.106420924406472	0.00183431375548407\\
0.107902879151618	0.00286463523699237\\
0.109405470720574	0.00412247212944898\\
0.110928986489522	0.00560696015707477\\
0.112473717836475	0.00731708026424457\\
0.114039960197003	0.00925165977568244\\
0.115628013120738	0.0114093737304279\\
0.11723818032866	0.0137887463874024\\
0.11887076977119	0.0163881529001357\\
0.120526093687084	0.0192058211579391\\
0.122204468663149	0.0222398337905508\\
0.123906215694792	0.0254881303330235\\
0.125631660247412	0.0289485095473675\\
0.127381132318648	0.032618631897228\\
0.129154966501488	0.0364960221716311\\
0.130953502048267	0.040578072253607\\
0.132777082935543	0.0448620440292826\\
0.134626057929891	0.0493450724328181\\
0.136500780654601	0.0540241686223663\\
0.138401609657313	0.0588962232820382\\
0.140328908478587	0.0639580100446736\\
0.142283045721435	0.0692061890300494\\
0.144264395121816	0.0746373104929913\\
0.146273335620113	0.0802478185757055\\
0.14831025143361	0.0860340551585099\\
0.150375532129974	0.0919922638030089\\
0.152469572701757	0.0981185937816518\\
0.154592773641948	0.104409104187491\\
0.156745541020559	0.110859768117886\\
0.158928286562298	0.117466476925797\\
0.161141427725302	0.124225044532249\\
0.163385387780986	0.131131211793511\\
0.165660595894991	0.138180650916444\\
0.167967487209265	0.145368969915487\\
0.170306502925284	0.152691717104709\\
0.172678090388436	0.160144385618326\\
0.175082703173572	0.167722417953131\\
0.177520801171763	0.175421210526265\\
0.179992850678248	0.183236118241787\\
0.182499324481615	0.191162459059557\\
0.18504070195423	0.199195518559989\\
0.187617469143912	0.20733055449828\\
0.190230118866894	0.21556280134181\\
0.192879150802078	0.22388747478449\\
0.195565071586595	0.232299776231911\\
0.198288394912707	0.240794897251279\\
0.20104964162605	0.249368023980213\\
0.203849339825246	0.258014341488614\\
0.206688024962908	0.266729038087944\\
0.209566239948043	0.275507309582409\\
0.212484535249888	0.284344363456666\\
0.215443469003188	0.293235422994852\\
0.218443607114943	0.302175731325891\\
0.221485523372636	0.311160555390214\\
0.224569799553977	0.320185189823194\\
0.227697025538168	0.329244960750804\\
0.230867799418717	0.338335229493181\\
0.234082727617829	0.347451396171995\\
0.237342425002387	0.356588903217688\\
0.240647515001542	0.365743238772921\\
0.243998629725955	0.37490993998871\\
0.247396410088681	0.384084596209979\\
0.250841505927754	0.393262852047494\\
0.254334576130465	0.402440410333317\\
0.25787628875938	0.411613034957191\\
0.261467321180109	0.420776553581456\\
0.265108360190854	0.429926860232333\\
0.268800102153761	0.439059917765661\\
0.272543253128103	0.448171760205347\\
0.276338529005317	0.457258494953092\\
0.28018665564592	0.466316304868094\\
0.28408836901833	0.475341450215735\\
0.28804441533963	0.484330270484433\\
0.292055551218275	0.493279186070068\\
0.296122543798803	0.502184699827638\\
0.300246170908555	0.511043398489978\\
0.30442722120643	0.51985195395363\\
0.308666494333727	0.528607124432115\\
0.312964801067075	0.537305755477124\\
0.317322963473498	0.545944780868277\\
0.321741815067637	0.554521223372367\\
0.326222200971167	0.563032195373144\\
0.330764978074424	0.571474899372906\\
0.335371015200293	0.579846628367364\\
0.340041193270371	0.588144766095371\\
0.344776405473446	0.596366787165341\\
0.349577557436328	0.604510257060303\\
0.354445567397043	0.612572832023707\\
0.359381366380463	0.620552258828268\\
0.364385898376354	0.628446374430241\\
0.36946012051993	0.636253105511713\\
0.374605003274899	0.643970467913569\\
0.379821530619073	0.651596565961968\\
0.385110700232557	0.659129591691253\\
0.390473523688556	0.666567823966329\\
0.395911026646846	0.673909627507668\\
0.401424249049932	0.681153451822177\\
0.407014245321944	0.688297830043252\\
0.412682084570295	0.695341377683452\\
0.418428850790158	0.702282791303255\\
0.424255643071778	0.709120847099477\\
0.430163575810679	0.715854399416951\\
0.4361537789208	0.722482379187142\\
0.44222739805059	0.729003792297423\\
0.448385594802119	0.735417717894744\\
0.45462954695324	0.74172330662751\\
0.460960448682843	0.747919778829459\\
0.467379510799246	0.75400642264939\\
0.473887960971765	0.759982592130569\\
0.480487043965513	0.765847705243688\\
0.487178021879463	0.771601241877214\\
0.493962174387832	0.777242741788988\\
0.500840798984821	0.782771802522897\\
0.507815211232767	0.788188077294463\\
0.51488674501375	0.79349127284913\\
0.522056752784698	0.798681147297034\\
0.529326605836056	0.803757507927989\\
0.536697694554048	0.808720209010403\\
0.544171428686589	0.813569149577787\\
0.551749237612913	0.81830427120647\\
0.559432570616938	0.822925555788102\\
0.567222897164454	0.827433023300453\\
0.575121707184161	0.83182672957998\\
0.583130511352622	0.83610676409955\\
0.591250841383188	0.840273247754681\\
0.599484250318941	0.844326330661546\\
0.607832312829723	0.848266189969972\\
0.616296625513294	0.852093027694552\\
0.624878807200689	0.855807068566938\\
0.633580499265825	0.859408557912305\\
0.642403365939419	0.862897759552883\\
0.65134909462728	0.866274953741404\\
0.66041939623303	0.869540435127207\\
0.669616005485322	0.872694510757669\\
0.678940681269611	0.875737498117556\\
0.68839520696455	0.878669723208794\\
0.697981390783066	0.881491518673081\\
0.707701066118189	0.884203221959676\\
0.717556091893693	0.88680517354062\\
0.727548352919623	0.889297715175548\\
0.737679760252773	0.891681188228171\\
0.747952251562182	0.893955932036428\\
0.758367791499719	0.896122282338219\\
0.768928372075831	0.898180569754522\\
0.779636013040523	0.900131118331659\\
0.790492762269642	0.901974244144341\\
0.801500696156541	0.903710253961067\\
0.812661920009195	0.905339443973359\\
0.823978568452852	0.906862098590238\\
0.835452805838287	0.908278489299241\\
0.847086826655741	0.909588873595223\\
0.858882855954625	0.910793493978091\\
0.870843149769072	0.91189257702053\\
0.882969995549409	0.912886332506717\\
0.89526571259964	0.913774952642919\\
0.907732652521023	0.914558611340802\\
0.920373199661822	0.915237463574198\\
0.933189771573324	0.915811644809981\\
0.9461848194722	0.916281270513659\\
0.959360828709314	0.916646435730167\\
0.972720319245054	0.916907214740299\\
0.986265846131282	0.917063660793126\\
1	0.917115805914674\\
1.01392540755882	0.917063660793052\\
1.02804473209331	0.916907214740151\\
1.0423606739764	0.916646435729944\\
1.0568759711848	0.916281270513357\\
1.07159339982267	0.915811644809596\\
1.08651577465254	0.915237463573726\\
1.10164594963366	0.914558611340238\\
1.11698681846782	0.913774952642255\\
1.13254131515281	0.912886332505946\\
1.14831241454351	0.911892577019642\\
1.16430313292088	0.910793493977076\\
1.18051652856881	0.909588873594069\\
1.19695570235904	0.908278489297934\\
1.21362379834424	0.906862098588763\\
1.23052400435926	0.905339443971698\\
1.24765955263087	0.903710253959202\\
1.2650337203959	0.901974244142251\\
1.28264983052806	0.900131118329319\\
1.30051125217341	0.898180569751905\\
1.31862140139475	0.896122282335295\\
1.33698374182495	0.893955932033166\\
1.35560178532937	0.891681188224533\\
1.37447909267754	0.889297715171495\\
1.39361927422414	0.886805173536107\\
1.41302599059953	0.884203221954652\\
1.43270295340983	0.881491518667493\\
1.45265392594678	0.878669723202581\\
1.4728827239075	0.875737498110652\\
1.49339321612425	0.87269451075\\
1.51418932530435	0.869540435118692\\
1.53527502878042	0.866274953731954\\
1.55665435927106	0.862897759542399\\
1.57833140565212	0.85940855790068\\
1.6003103137387	0.855807068554052\\
1.62259528707809	0.852093027680273\\
1.64519058775366	0.848266189954158\\
1.66810053720006	0.844326330644038\\
1.69132951702965	0.840273247735305\\
1.71488196987054	0.836106764078115\\
1.73876240021625	0.831826729556277\\
1.76297537528721	0.827433023274252\\
1.78752552590424	0.822925555759152\\
1.81241754737424	0.818304271174495\\
1.83765620038817	0.813569149542486\\
1.86324631193156	0.808720208971445\\
1.88919277620767	0.803757507885012\\
1.91550055557353	0.798681147249645\\
1.94217468148903	0.793491272796897\\
1.96922025547917	0.788188077236914\\
1.99664245010979	0.782771802459518\\
2.0244465099768	0.777242741719217\\
2.05263775270925	0.771601241800441\\
2.08122156998634	0.765847705159244\\
2.11020342856859	0.75998259203773\\
2.13958887134342	0.754006422547365\\
2.16938351838518	0.747919778717389\\
2.19959306803007	0.74172330650446\\
2.23022329796594	0.735417717759699\\
2.26128006633728	0.729003792149281\\
2.29276931286565	0.722482379024707\\
2.32469705998565	0.715854399238924\\
2.35706941399673	0.709120846904454\\
2.38989256623105	0.702282791089713\\
2.4231727942376	0.695341377449742\\
2.45691646298279	0.688297829787594\\
2.49113002606779	0.681153451542642\\
2.52582002696278	0.673909627202176\\
2.56099310025846	0.666567823632632\\
2.59665597293487	0.65912959132693\\
2.63281546564802	0.651596565564409\\
2.66947849403432	0.64397046747996\\
2.70665207003324	0.636253105039029\\
2.74434330322837	0.628446373915228\\
2.78255940220713	0.620552258267428\\
2.82130767593947	0.612572831413289\\
2.86059553517574	0.604510256396278\\
2.90043049386399	0.596366786443396\\
2.94082017058706	0.588144765310883\\
2.98177229001967	0.57984662751539\\
3.02329468440578	0.571474898448161\\
3.06539529505653	0.563032194369984\\
3.10808217386906	0.554521222284772\\
3.15136348486648	0.545944779689829\\
3.19524750575921	0.53730575420099\\
3.2397426295282	0.528607123051029\\
3.28485736603004	0.519851952459868\\
3.33060034362459	0.511043396875347\\
3.37698031082509	0.50218469808345\\
3.42400613797143	0.493279184187125\\
3.47168681892656	0.484330268453008\\
3.52003147279668	0.475341448025553\\
3.56904934567523	0.466316302508318\\
3.61874981241128	0.457258492412308\\
3.66914237840249	0.448171757471547\\
3.72023668141307	0.439059914826235\\
3.772042493417	0.429926857074057\\
3.824569722467	0.420776550190482\\
3.87782841458946	0.41161303131905\\
3.93182875570577	0.40244040643291\\
3.98658107358044	0.3932628478691\\
4.04209583979631	0.384084591737257\\
4.09838367175726	0.374909935204714\\
4.15545533471888	0.365743233660114\\
4.21332174384729	0.356588897757966\\
4.27199396630678	0.347451390346718\\
4.3314832233764	0.338335223283206\\
4.39180089259609	0.329244954136528\\
4.45295850994266	0.320185182784611\\
4.5149677720361	0.311160547906974\\
4.57784053837662	0.302175723377372\\
4.64158883361278	0.293235414560244\\
4.70622484984128	0.284344354515065\\
4.77176094893875	0.275507300112926\\
4.83820966492596	0.266729028069824\\
4.90558370636505	0.25801433090137\\
4.97389595879006	0.249368012803779\\
5.04315948717136	0.24079488546617\\
5.11338753841433	0.232299763819411\\
5.18459354389291	0.223887461726849\\
5.25679112201842	0.21556278762246\\
5.32999408084409	0.207330540102072\\
5.40421642070591	0.199195503473444\\
5.47947233690029	0.191162443271138\\
5.55577622239888	0.183236101742195\\
5.63314267060136	0.175421193308743\\
5.71158647812643	0.167722400013792\\
5.79112264764176	0.1601443669565\\
5.87176639073326	0.152691697723302\\
5.95353313081437	0.145368949821354\\
6.03643850607587	0.138180630120775\\
6.1204983724767	0.131131190312233\\
6.2057288067765	0.12422502238642\\
6.29214610961034	0.117466454142016\\
6.37976680860628	0.110859744728695\\
6.46860766154633	0.104409080231766\\
6.55868565957143	0.0981185693049825\\
6.65001803043112	0.0919922388580561\\
6.74262224177834	0.0860340298053269\\
6.83651600451024	0.0802477928820221\\
6.93171727615541	0.0746372845344381\\
7.02824426430835	0.0692061628903177\\
7.12611543011175	0.0639579838155874\\
7.22534949178721	0.0588961970635276\\
7.32596542821523	0.0540241425223158\\
7.42798248256492	0.0493450465667712\\
7.53142016597438	0.0448620185199753\\
7.63629826128224	0.0405780472303012\\
7.74263682681127	0.0364959977692162\\
7.85045620020451	0.0326186082550576\\
7.95977700231499	0.0289484868077859\\
8.07062014114951	0.0254881086395435\\
8.18300681586739	0.0222398132856301\\
8.29695852083491	0.0192058019803025\\
8.41249704973612	0.0163881351815855\\
8.52964449974102	0.0137887302490474\\
8.64842327573173	0.0114093592782627\\
8.76885609458743	0.00925164709543515\\
8.89096598952916	0.00731706941540763\\
9.01477631452492	0.00560695116602409\\
9.14031074875623	0.00412246498154598\\
9.26759330114688	0.00286462986755803\\
9.39664831495469	0.00183431003952073\\
9.52750047242729	0.00103221393685074\\
9.66017479952265	0.000458893414128654\\
9.7946966706954	0.000114743110743486\\
9.9310918137498	9.99994029065213e-16\\
10.0693863147603	0.000114743118422447\\
10.2096066230605	0.000458893475694206\\
10.3517795563018	0.00103221414538328\\
10.4959323055823	0.00183431053630898\\
10.6420924406472	0.00286463084411183\\
10.7902879151618	0.00412246668229666\\
10.9405470720574	0.00560695389173069\\
11.0928986489522	0.0073170735272853\\
11.2473717836475	0.00925165302003044\\
11.4039960197003	0.0114093675131031\\
11.5628013120738	0.0137887413690954\\
11.723818032866	0.0163881498465328\\
11.887076977119	0.0192058209427432\\
12.0526093687084	0.0222398374001564\\
12.2204468663149	0.0254881388728111\\
12.3906215694792	0.0289485242495996\\
12.5631660247412	0.0326186541305365\\
12.7381132318648	0.0364960534520927\\
12.9154966501488	0.0405781142574227\\
13.0953502048267	0.0448620986070768\\
13.2777082935543	0.0493451416255875\\
13.4626057929891	0.0540242546791186\\
13.6500780654601	0.0588963286791619\\
13.8401609657313	0.0639581375070981\\
14.0328908478587	0.0692063415542563\\
14.2283045721435	0.0746374913719467\\
14.4264395121816	0.0802480314257907\\
14.6273335620113	0.0860343039485377\\
14.831025143361	0.0919925528854149\\
15.0375532129974	0.0981189279259589\\
15.2469572701757	0.104409488616159\\
15.4592773641948	0.110860208544647\\
15.674554102056	0.11746697959661\\
15.8928286562298	0.124225616268983\\
16.1141427725302	0.131131860040495\\
16.3385387780986	0.138181383790023\\
16.5660595894991	0.14536979625672\\
16.7967487209265	0.152692646535354\\
17.0306502925284	0.160145428600278\\
17.2678090388436	0.167723585851457\\
17.5082703173573	0.175422515676\\
17.7520801171763	0.18323757401867\\
17.9992850678248	0.191164079954868\\
18.2499324481615	0.199197320259661\\
18.504070195423	0.207332553966483\\
18.7617469143912	0.215565016909169\\
19.0230118866894	0.223889926241141\\
19.2879150802078	0.232302484925583\\
19.5565071586595	0.240797886190591\\
19.8288394912707	0.249371317943384\\
20.104964162605	0.258017967137792\\
20.3849339825246	0.26673302408935\\
20.6688024962908	0.275511686732498\\
20.9566239948043	0.284349164814513\\
21.2484535249888	0.293240684020967\\
21.5443469003188	0.30218149002767\\
21.8443607114943	0.31116685247423\\
22.1485523372636	0.32019206885454\\
22.4569799553977	0.329252468319686\\
22.7697025538168	0.33834341538897\\
23.0867799418717	0.347460313564932\\
23.4082727617829	0.356598608848463\\
23.7342425002387	0.365753793150297\\
24.0647515001542	0.3749214075954\\
24.3998629725955	0.384097045716971\\
24.7396410088681	0.393276356536993\\
25.0841505927754	0.402455047530502\\
25.4334576130465	0.411628887470949\\
25.787628875938	0.420793709154267\\
26.1467321180109	0.429945411999469\\
26.5108360190854	0.439079964523841\\
26.8800102153761	0.448193406691009\\
27.2543253128103	0.457281852130403\\
27.6338529005317	0.466341490226839\\
28.018665564592	0.475368588079182\\
28.408836901833	0.484359492327291\\
28.804441533963	0.493310630846633\\
29.2055551218275	0.502218514310184\\
29.6122543798803	0.511079737617494\\
30.0246170908555	0.519890981190915\\
30.442722120643	0.5286490121393\\
30.8666494333727	0.53735068528962\\
31.2964801067075	0.545992944087163\\
31.7322963473498	0.554572821365204\\
32.1741815067637	0.563087439985173\\
32.6222200971167	0.571534013348601\\
33.0764978074424	0.579909845782241\\
33.5371015200293	0.588212332797992\\
34.0041193270371	0.596438961229379\\
34.4776405473446	0.604587309246548\\
34.9577557436328	0.612655046251851\\
35.4445567397044	0.620639932658293\\
35.9381366380463	0.628539819553211\\
36.4385898376354	0.63635264824973\\
36.946012051993	0.644076449728659\\
37.4605003274899	0.651709343973607\\
37.9821530619074	0.659249539202229\\
38.5110700232557	0.666695330996618\\
39.0473523688556	0.674045101335956\\
39.5911026646846	0.681297317534641\\
40.1424249049932	0.688450531089198\\
40.7014245321944	0.695503376437348\\
41.2682084570295	0.702454569632698\\
41.8428850790158	0.709302906938588\\
42.4255643071778	0.716047263344656\\
43.016357581068	0.722686591009788\\
43.6153778920801	0.729219917635121\\
44.222739805059	0.735646344770818\\
44.8385594802119	0.741965046060394\\
45.462954695324	0.748175265426347\\
46.0960448682843	0.754276315200915\\
46.7379510799246	0.760267574205767\\
47.3887960971766	0.766148485784441\\
48.0487043965513	0.771918555791371\\
48.7178021879463	0.777577350541292\\
49.3962174387832	0.783124494722848\\
50.0840798984821	0.788559669280189\\
50.7815211232768	0.793882609266316\\
51.4886745013749	0.799093101671927\\
52.2056752784698	0.804190983233464\\
52.9326605836056	0.809176138224037\\
53.6697694554048	0.814048496230859\\
54.4171428686589	0.818808029922784\\
55.1749237612912	0.823454752811475\\
55.9432570616938	0.827988717009705\\
56.7222897164454	0.832410010990214\\
57.5121707184161	0.836718757348487\\
58.3130511352622	0.840915110572777\\
59.1250841383188	0.844999254824616\\
59.9484250318941	0.848971401732972\\
60.7832312829723	0.852831788205188\\
61.6296625513294	0.856580674257715\\
62.4878807200689	0.86021834086961\\
63.3580499265825	0.86374508786168\\
64.2403365939419	0.867161231804087\\
65.134909462728	0.870467103955123\\
66.0419396233031	0.873663048233832\\
66.9616005485322	0.876749419229018\\
67.8940681269611	0.879726580247144\\
68.839520696455	0.882594901401509\\
69.7981390783066	0.88535475774504\\
70.7701066118189	0.888006527448919\\
71.7556091893693	0.890550590029204\\
72.7548352919623	0.892987324623518\\
73.7679760252773	0.895317108319779\\
74.7952251562182	0.897540314538878\\
75.8367791499719	0.899657311473128\\
76.8928372075831	0.901668460582206\\
77.9636013040524	0.903574115148248\\
79.0492762269642	0.905374618891667\\
80.150069615654	0.907070304649171\\
81.2661920009195	0.9086614931154\\
82.3978568452852	0.910148491649494\\
83.5452805838287	0.911531593147848\\
84.708682665574	0.912811074984212\\
85.8882855954625	0.91398719801822\\
87.0843149769072	0.915060205673367\\
88.2969995549409	0.916030323085347\\
89.526571259964	0.916897756321628\\
90.7732652521022	0.917662691673014\\
92.0373199661822	0.91832529501792\\
93.3189771573324	0.918885711259967\\
94.61848194722	0.919344063839455\\
95.9360828709315	0.919700454319196\\
97.2720319245054	0.919954962045096\\
98.6265846131282	0.92010764388183\\
100	0.920158534023849\\
101.392540755881	0.920107643881913\\
102.804473209331	0.919954962045262\\
104.23606739764	0.919700454319446\\
105.687597118481	0.919344063839794\\
107.159339982267	0.918885711260399\\
108.651577465254	0.91832529501845\\
110.164594963366	0.917662691673648\\
111.698681846782	0.916897756322373\\
113.254131515281	0.916030323086213\\
114.831241454351	0.915060205674363\\
116.430313292088	0.91398719801936\\
118.051652856881	0.912811074985507\\
119.695570235904	0.911531593149314\\
121.362379834424	0.910148491651148\\
123.052400435926	0.908661493117261\\
124.765955263087	0.907070304651262\\
126.50337203959	0.905374618894011\\
128.264983052806	0.903574115150872\\
130.051125217341	0.901668460585139\\
131.862140139475	0.899657311476404\\
133.698374182495	0.897540314542533\\
135.560178532937	0.895317108323854\\
137.447909267754	0.892987324628058\\
139.361927422414	0.890550590034259\\
141.302599059953	0.888006527454544\\
143.270295340983	0.885354757751297\\
145.265392594678	0.882594901408465\\
147.28827239075	0.879726580254872\\
149.339321612425	0.876749419237602\\
151.418932530435	0.87366304824336\\
153.527502878042	0.870467103965697\\
155.665435927106	0.867161231815815\\
157.833140565212	0.863745087874685\\
160.03103137387	0.860218340884023\\
162.259528707809	0.856580674273683\\
164.519058775366	0.852831788222872\\
166.810053720006	0.848971401752548\\
169.132951702965	0.844999254846278\\
171.488196987054	0.840915110596738\\
173.876240021625	0.836718757374981\\
176.29753752872	0.832410011019498\\
178.752552590424	0.827988717042058\\
181.241754737424	0.823454752847204\\
183.765620038817	0.818808029962227\\
186.324631193156	0.814048496274385\\
188.919277620767	0.809176138272048\\
191.550055557353	0.804190983286402\\
194.217468148902	0.799093101730273\\
196.922025547917	0.793882609330595\\
199.664245010979	0.788559669350976\\
202.444650997681	0.783124494800768\\
205.263775270925	0.777577350627029\\
208.122156998634	0.771918555885669\\
211.02034285686	0.766148485888109\\
213.958887134342	0.760267574319687\\
216.938351838519	0.754276315326048\\
219.959306803007	0.748175265563735\\
223.022329796594	0.741965046211171\\
226.128006633728	0.735646344936213\\
229.276931286565	0.729219917816471\\
232.469705998565	0.722686591208542\\
235.706941399673	0.716047263562383\\
238.989256623105	0.70930290717699\\
242.31727942376	0.702454569893615\\
245.691646298279	0.695503376722772\\
249.113002606779	0.688450531401283\\
252.582002696278	0.681297317875712\\
256.099310025846	0.674045101708523\\
259.665597293487	0.66669533140339\\
263.281546564802	0.659249539646126\\
266.947849403432	0.651709344457773\\
270.665207003324	0.644076450256481\\
274.434330322836	0.63635264882485\\
278.255940220713	0.628539820179541\\
282.130767593947	0.620639933340036\\
286.059553517574	0.612655046993514\\
290.043049386399	0.604587310052967\\
294.082017058706	0.596438962105733\\
298.177229001967	0.588212333749821\\
302.329468440578	0.579909846815472\\
306.539529505653	0.571534014469564\\
310.808217386906	0.563087441200625\\
315.136348486648	0.554572822682349\\
319.524750575921	0.545992945513674\\
323.97426295282	0.537350686833663\\
328.485736603005	0.528649013809555\\
333.060034362459	0.5198909829966\\
337.698031082509	0.511079739568386\\
342.400613797143	0.502218516416638\\
347.168681892656	0.493310633119609\\
352.003147279668	0.484359494778374\\
356.904934567523	0.475368590720596\\
361.874981241128	0.466341493071473\\
366.914237840249	0.457281855191822\\
372.023668141307	0.448193409983474\\
377.2042493417	0.439079968062318\\
382.4569722467	0.429945415799646\\
387.782841458945	0.420793713232553\\
393.182875570577	0.411628891844481\\
398.658107358044	0.402455052217151\\
404.209583979631	0.393276361555351\\
409.838367175726	0.38409705108635\\
415.545533471887	0.374921413335811\\
421.332174384729	0.365753799282427\\
427.199396630678	0.356598615393654\\
433.14832233764	0.34746032054514\\
439.180089259609	0.338343422826723\\
445.295850994265	0.329252476238029\\
451.49677720361	0.320192077276967\\
457.784053837662	0.311166861424618\\
464.158883361278	0.302181499530184\\
470.622484984128	0.293240694099965\\
477.176094893874	0.284349175494432\\
483.820966492596	0.275511698037721\\
490.558370636505	0.266733036044069\\
497.389595879007	0.258017979765842\\
504.315948717136	0.24937133126807\\
511.338753841432	0.240797900234488\\
518.459354389291	0.23230249971032\\
525.679112201842	0.223889941787168\\
532.999408084409	0.215565033235498\\
540.421642070591	0.207332571090417\\
547.947233690028	0.199197338196491\\
555.577622239888	0.191164098717549\\
563.314267060135	0.18323759361749\\
571.158647812643	0.175422536118208\\
579.112264764176	0.167723607140886\\
587.176639073326	0.160145450736942\\
595.353313081437	0.152692669515027\\
603.643850607586	0.145369820070498\\
612.04983724767	0.138181408423879\\
620.57288067765	0.131131885474817\\
629.214610961035	0.124225642478105\\
637.976680860628	0.117467006548353\\
646.860766154632	0.110860236199848\\
655.868565957144	0.104409516928219\\
665.001803043112	0.0981189568404146\\
674.262224177835	0.0919925823395272\\
683.651600451024	0.0860343338709295\\
693.17172761554	0.0802480617361267\\
702.824426430835	0.0746375219806848\\
712.611543011174	0.06920637236248\\
722.534949178722	0.0639581684064429\\
732.596542821523	0.0588963595518695\\
742.798248256491	0.0540242853982298\\
753.142016597438	0.0493451720553066\\
763.629826128224	0.0448621286033396\\
774.263682681128	0.0405781436686925\\
785.045620020451	0.03649608212043\\
795.977700231498	0.0326186818929781\\
807.062014114951	0.0289485509398958\\
818.300681586739	0.0254881643235717\\
829.695852083491	0.0222398614454593\\
841.249704973612	0.019205843421271\\
852.964449974102	0.0163881706053016\\
864.842327573173	0.0137887602678547\\
876.885609458743	0.0114093844294888\\
889.096598952917	0.00925166785555845\\
901.477631452492	0.0073170862142899\\
914.031074875622	0.00560696440135058\\
926.759330114688	0.00412247503363052\\
939.664831495469	0.0028646371146723\\
952.75004724273	0.00183431487391295\\
966.017479952265	0.00103221678163283\\
979.469667069539	0.000458894741213132\\
993.10918137498	0.000114743460027054\\
1006.93863147603	9.99997242245985e-16\\
1020.96066230605	0.000114743508580688\\
1035.17795563018	0.000458895130581805\\
1049.59323055823	0.00103221810105039\\
1064.20924406472	0.00183431801903756\\
1079.02879151618	0.00286464330184891\\
1094.05470720574	0.00412248581906082\\
1109.28986489522	0.00560698170530259\\
1124.73717836475	0.00731711235051653\\
1140.39960197003	0.00925170556612146\\
1156.28013120738	0.0114094369252252\\
1172.3818032866	0.013788831274759\\
1188.7076977119	0.0163882644171224\\
1205.26093687084	0.0192059649586759\\
1222.04468663149	0.0222400163221386\\
1239.06215694792	0.0254883589197044\\
1256.31660247412	0.0289487924834403\\
1273.81132318648	0.032618978549274\\
1291.54966501488	0.0364964430906677\\
1309.53502048267	0.0405785792978185\\
1327.77082935543	0.044862650498031\\
1346.26057929891	0.0493457932126905\\
1365.00780654601	0.0540250203460531\\
1384.01609657313	0.0588972245009019\\
1403.28908478587	0.0639591814159135\\
1422.83045721435	0.0692075535194182\\
1442.64395121816	0.0746388935940932\\
1462.73335620113	0.0802496485469392\\
1483.1025143361	0.0860361632788013\\
1503.75532129974	0.0919946846475282\\
1524.69572701757	0.098121365518768\\
1545.92773641948	0.104412268898302\\
1567.45541020559	0.1108633721397\\
1589.28286562298	0.117470571221034\\
1611.41427725302	0.124229685084289\\
1633.85387780986	0.131136460031077\\
1656.60595894992	0.13818657416821\\
1679.67487209265	0.145375641896637\\
1703.06502925284	0.152699218437279\\
1726.78090388436	0.160152804387229\\
1750.82703173572	0.167731850299832\\
1775.20801171764	0.175431761282162\\
1799.92850678248	0.183247901603443\\
1824.99324481615	0.191175599307997\\
1850.4070195423	0.199210150826351\\
1876.17469143912	0.207346825578224\\
1902.30118866895	0.215580870561138\\
1928.79150802078	0.223907514918534\\
1955.65071586595	0.232321974481334\\
1982.88394912707	0.240819456277001\\
2010.4964162605	0.24939516300029\\
2038.49339825246	0.258044297439947\\
2066.88024962908	0.26676206685582\\
2095.66239948043	0.27554368730093\\
2124.84535249888	0.284384387883219\\
2154.43469003188	0.29327941496187\\
2184.43607114943	0.302224036273221\\
2214.85523372636	0.311213544981484\\
2245.69799553978	0.320243263649697\\
2276.97025538168	0.329308548126428\\
2308.67799418717	0.338404791344066\\
2340.82727617829	0.347527427024603\\
2373.42425002387	0.356671933289117\\
2406.47515001543	0.365833836167304\\
2439.98629725955	0.375008713003642\\
2473.96410088681	0.384192195756992\\
2508.41505927754	0.393379974190626\\
2543.34576130465	0.402567798949908\\
2578.7628875938	0.411751484525102\\
2614.67321180109	0.420926912096922\\
2651.08360190854	0.430090032262765\\
2688.00102153761	0.439236867641698\\
2725.43253128103	0.448363515356543\\
2763.38529005317	0.457466149391637\\
2801.8665564592	0.466541022824992\\
2840.8836901833	0.475584469933914\\
2880.4441533963	0.484592908173241\\
2920.55551218275	0.493562840025651\\
2961.22543798804	0.502490854723701\\
3002.46170908555	0.511373629843405\\
3044.2722120643	0.520207932769463\\
3086.66494333727	0.528990622032363\\
3129.64801067075	0.53771864851784\\
3173.22963473498	0.54638905654935\\
3217.41815067637	0.554998984844395\\
3262.22200971167	0.56354566734574\\
3307.64978074424	0.572026433928739\\
3353.71015200293	0.580438710986143\\
3400.41193270371	0.588780021891957\\
3447.76405473446	0.597047987346068\\
3495.77557436327	0.605240325601491\\
3544.45567397044	0.6133548525763\\
3593.81366380463	0.62138948185236\\
3643.85898376355	0.62934222456321\\
3694.6012051993	0.637211189173491\\
3746.05003274899	0.644994581152494\\
3798.21530619074	0.652690702544496\\
3851.10700232557	0.660297951438637\\
3904.73523688556	0.667814821341257\\
3959.11026646846	0.675239900453615\\
4014.24249049932	0.682571870858085\\
4070.14245321944	0.689809507615935\\
4126.82084570295	0.696951677779905\\
4184.28850790159	0.703997339324858\\
4242.55643071778	0.710945539999799\\
4301.63575810679	0.717795416104683\\
4361.53778920801	0.724546191195368\\
4422.2739805059	0.731197174720202\\
4483.85594802119	0.737747760591713\\
4546.2954695324	0.744197425696869\\
4609.60448682843	0.750545728349455\\
4673.79510799246	0.756792306688067\\
4738.87960971765	0.762936877023212\\
4804.87043965513	0.768979232137082\\
4871.78021879463	0.77491923953943\\
4939.62174387832	0.780756839683097\\
5008.40798984821	0.786492044142573\\
5078.15211232767	0.792124933759081\\
5148.8674501375	0.797655656755532\\
5220.56752784698	0.803084426824706\\
5293.26605836056	0.808411521193988\\
5366.97694554048	0.813637278669873\\
5441.71428686589	0.818762097665465\\
5517.49237612913	0.823786434214108\\
5594.32570616938	0.8287107999722\\
5672.22897164455	0.83353576021424\\
5751.21707184161	0.838261931822995\\
5831.30511352622	0.842889981277704\\
5912.50841383188	0.847420622643055\\
5994.84250318941	0.851854615561681\\
6078.32312829724	0.85619276325278\\
6162.96625513294	0.860435910519394\\
6248.78807200689	0.864584941766812\\
6335.80499265825	0.868640779034449\\
6424.03365939419	0.872604380043475\\
6513.49094627281	0.876476736262379\\
6604.19396233031	0.880258870992543\\
6696.16005485322	0.883951837475832\\
6789.40681269611	0.887556717026079\\
6883.9520696455	0.891074617186282\\
6979.81390783067	0.894506669913211\\
7077.01066118189	0.897854029791032\\
7175.56091893692	0.901117872275461\\
7275.48352919623	0.904299391969875\\
7376.79760252773	0.907399800934683\\
7479.52251562183	0.910420327031202\\
7583.67791499719	0.913362212301165\\
7689.28372075831	0.916226711382889\\
7796.36013040524	0.919015089965082\\
7904.92762269642	0.921728623279109\\
8015.00696156541	0.924368594630513\\
8126.61920009195	0.926936293970472\\
8239.78568452851	0.929433016507766\\
8354.52805838287	0.931860061361791\\
8470.8682665574	0.934218730257033\\
8588.82855954626	0.936510326259363\\
8708.43149769072	0.938736152554412\\
8829.69995549408	0.940897511268237\\
8952.6571259964	0.942995702330385\\
9077.32652521022	0.945032022379413\\
9203.73199661823	0.947007763710857\\
9331.89771573324	0.948924213267543\\
9461.84819472199	0.95078265167211\\
9593.60828709315	0.952584352301532\\
9727.20319245054	0.954330580403368\\
9862.65846131283	0.956022592253417\\
10000	0.957661634354408\\
};
\end{axis}
\end{tikzpicture}%

%% file: Figures/Exm_fom_H2norm.tikz
%
%
\begin{tikzpicture}

\begin{axis}[%
width=0.951\fwidth,
height=\fheight,
at={(0\fwidth,0\fheight)},
scale only axis,
xmin=10,
xmax=50,
ymode=log,
ymin=1e-15,
ymax=100000,
yminorticks=true,
xlabel = {Number of interpolation points},
ylabel = {$\|\bH - \widehat{\bH} \|_{\cH_{2,w}}$},
axis background/.style={fill=white},
legend style={font=\tiny,at={(0.990,0.18)},legend cell align=left, align=left, draw=white!15!black}
]
\addplot [color=green!50!black,dashed, line width = 2pt]
  table[row sep=crcr]{%
10	21322.3764492342\\
14	102.036728897932\\
18	1.11068860237512e-07\\
22	3.36936126174967e-09\\
26	2.23035583035997e-09\\
30	1.2826224584481e-09\\
};

\addplot [color=magenta,dashdotted, line width = 2pt]
  table[row sep=crcr]{%
10	31361.015005358\\
14	26197.6170687441\\
18	6197.25584747528\\
22	13.4468957552319\\
26	0.0739917001320123\\
30	7.73544863488795e-06\\
34	3.23534652367508e-08\\
38	1.95561579083292e-11\\
42	4.43232789802281e-09\\
46	2.02314285391808e-11\\
50	8.27308741539669e-10\\
};

\end{axis}
\end{tikzpicture}%

%% file: Figures/Exm_beam_H2norm.tikz
%
%
\begin{tikzpicture}

\begin{axis}[%
width=0.951\fwidth,
height=\fheight,
at={(0\fwidth,0\fheight)},
scale only axis,
xmin=10,
xmax=60,
ymode=log,
ymin=1e-12,
ymax=100,
yminorticks=true,
xlabel = {Number of interpolation points},
ylabel = {$\|\bH - \widehat{\bH} \|_{\cH_{2,w}}$},
axis background/.style={fill=white},
legend style={font=\tiny,at={(1,1.01)},legend cell align=left, align=left, draw=white!15!black}
]
\addplot [color=green!50!black,dashed, line width = 2pt]
  table[row sep=crcr]{%
10	5.4159714705843\\
14	0.601613901752364\\
18	0.000421445272560973\\
22	3.5246993960643e-05\\
26	2.8100992998612e-11\\
};

\addplot [color=magenta,dashdotted, line width = 2pt]
  table[row sep=crcr]{%
10	25.9159166083791\\
14	3.02180945599167\\
18	0.0886164726937016\\
22	0.0511328207139634\\
26	0.00134797782292346\\
30	7.6217912605204e-06\\
34	0.000108936134034932\\
38	4.845813608439e-05\\
42	2.41417931401601e-06\\
46	8.35267456816271e-07\\
50	4.71961476287194e-08\\
54	2.3869776024915e-08\\
58	7.05829930574664e-10\\
};

\end{axis}
\end{tikzpicture}%

%% file: Figures/Exm_RLCSerkan_time_step.tikz
%
%
\begin{tikzpicture}

\begin{axis}[%
width=0.951\fwidth,
height=\fheight,
at={(0\fwidth,0\fheight)},
scale only axis,
xmin=0,
xmax=4,
ymin=-3,
ymax=4,
axis background/.style={fill=white},
legend style={legend cell align=left, align=left, draw=white!15!black}
]
\addplot [color=blue, line width = 2pt]
  table[row sep=crcr]{%
0	0\\
0.01	1.38286659144485\\
0.02	1.84762933875602\\
0.03	1.9366721106004\\
0.04	1.85456969176914\\
0.05	1.68157806802536\\
0.06	1.45281533456646\\
0.07	1.18748852014761\\
0.08	0.899521655792258\\
0.09	0.601246287509456\\
0.1	0.304498076425755\\
0.11	0.0207444846335102\\
0.12	-0.239141509700594\\
0.13	-0.465222284361778\\
0.14	-0.648851855794625\\
0.15	-0.783001552405356\\
0.16	-0.862530233683294\\
0.17	-0.884383813065886\\
0.18	-0.847714387806317\\
0.19	-0.753913481200193\\
0.2	-0.606557684679809\\
0.21	-0.411268604019793\\
0.22	-0.175492471502573\\
0.23	0.0917919977826864\\
0.24	0.38042606010797\\
0.25	0.679469851281502\\
0.26	0.977632031288906\\
0.27	1.26371321049916\\
0.28	1.52704563847317\\
0.29	1.75791176864616\\
0.3	1.94792513600344\\
0.31	2.09035846808901\\
0.32	2.18040603372283\\
0.33	2.21536983610074\\
0.34	2.19476227423667\\
0.35	2.12032120909699\\
0.36	1.99593684655204\\
0.37	1.82749335012597\\
0.38	1.622631483086\\
0.39	1.39044171678922\\
0.4	1.1411000053755\\
0.41	0.88546070570143\\
0.42	0.634622824944093\\
0.43	0.399486838550255\\
0.44	0.190319695756096\\
0.45	0.0163453036523756\\
0.46	-0.114623233466761\\
0.47	-0.1964950047281\\
0.48	-0.225143942244863\\
0.49	-0.198573094780798\\
0.5	-0.116994010466832\\
0.51	0.017181907253185\\
0.52	0.199439702157697\\
0.53	0.423340968502627\\
0.54	0.680780561379403\\
0.55	0.962309671186508\\
0.56	1.25751237780895\\
0.57	1.55542067187173\\
0.58	1.84495140018818\\
0.59	2.11534772123927\\
0.6	2.35660747892088\\
0.61	2.5598814259739\\
0.62	2.71782543163013\\
0.63	2.82489264314011\\
0.64	2.87755396488199\\
0.65	2.87443807629413\\
0.66	2.81638541706841\\
0.67	2.70641399708452\\
0.68	2.54959840281736\\
0.69	2.35286683136611\\
0.7	2.124724249978\\
0.71	1.87491272278865\\
0.72	1.61402245016199\\
0.73	1.35306902971066\\
0.74	1.10305379356307\\
0.75	0.874524750099278\\
0.76	0.677155633371055\\
0.77	0.519359840758548\\
0.78	0.407954647916856\\
0.79	0.347889085203795\\
0.8	0.342046321480214\\
0.81	0.391128430621244\\
0.82	0.493628131685463\\
0.83	0.645888626383016\\
0.84	0.842249145478568\\
0.85	1.07527039905256\\
0.86	1.33603094032413\\
0.87	1.61448262597536\\
0.88	1.89985100028758\\
0.89	2.18106463980251\\
0.9	2.44719634102494\\
0.91	2.68789856189146\\
0.92	2.89381575714095\\
0.93	3.05695716918246\\
0.94	3.17101521282387\\
0.95	3.23161676143462\\
0.96	3.2364973172889\\
0.97	3.18559112331788\\
0.98	3.08103362472547\\
0.99	2.92707618328943\\
1	2.72991644537497\\
1.01	2.49745112727047\\
1.02	2.23896107436853\\
1.03	1.96474115065628\\
1.04	1.68568971431818\\
1.05	1.41287404631436\\
1.06	1.15708905735719\\
1.07	0.928426866556457\\
1.08	0.735874411466873\\
1.09	0.586955131636819\\
1.1	0.487429010583972\\
1.11	0.441062934467591\\
1.12	0.449480522345981\\
1.13	0.512097414563845\\
1.14	0.626144598821709\\
1.15	0.786778843661016\\
1.16	0.987275836381728\\
1.17	1.21929832523823\\
1.18	1.47322857557933\\
1.19	1.73855188561852\\
1.2	2.00427587195643\\
1.21	2.2593688089782\\
1.22	2.49319954666281\\
1.23	2.69596146844996\\
1.24	2.85906358711795\\
1.25	2.97547318676532\\
1.26	3.03999635072525\\
1.27	3.04948519156443\\
1.28	3.00296352150074\\
1.29	2.90166595311909\\
1.3	2.74898887154354\\
1.31	2.55035523264199\\
1.32	2.31299857733256\\
1.33	2.04567487266201\\
1.34	1.7583136676438\\
1.35	1.46162247116723\\
1.36	1.16666012416861\\
1.37	0.88439617434456\\
1.38	0.625273819708139\\
1.39	0.398793844995147\\
1.4	0.21313613799251\\
1.41	0.0748338746510404\\
1.42	-0.0114866379091534\\
1.43	-0.0432900889816315\\
1.44	-0.0202329219938914\\
1.45	0.0558232943802142\\
1.46	0.180886414620133\\
1.47	0.348993212090654\\
1.48	0.552447463336833\\
1.49	0.782127129086638\\
1.5	1.02784838543388\\
1.51	1.27877199223346\\
1.52	1.52383579788737\\
1.53	1.75219613772442\\
1.54	1.95366052861694\\
1.55	2.11909440947113\\
1.56	2.24078571193805\\
1.57	2.3127527268712\\
1.58	2.33098299267872\\
1.59	2.29359368165956\\
1.6	2.20090709004412\\
1.61	2.05543822200876\\
1.62	1.86179496247116\\
1.63	1.62649481828508\\
1.64	1.35770553360776\\
1.65	1.06491992011008\\
1.66	0.758577865346971\\
1.67	0.449650588443284\\
1.68	0.149203717330787\\
1.69	-0.132043393912301\\
1.7	-0.38414453917517\\
1.71	-0.598322707941745\\
1.72	-0.767319517450692\\
1.73	-0.885684098820273\\
1.74	-0.94998992035223\\
1.75	-0.958970904150925\\
1.76	-0.913571409714132\\
1.77	-0.8169080920101\\
1.78	-0.674145154841182\\
1.79	-0.492287972954127\\
1.8	-0.279903310740259\\
1.81	-0.04677729173074\\
1.82	0.196475245225308\\
1.83	0.438834429153066\\
1.84	0.669315188390811\\
1.85	0.877405695372057\\
1.86	1.05348694512916\\
1.87	1.18921674003515\\
1.88	1.27786277231313\\
1.89	1.31457152515338\\
1.9	1.29656227198838\\
1.91	1.22323843957904\\
1.92	1.09621189500688\\
1.93	0.9192391881212\\
1.94	0.698072291045364\\
1.95	0.440229785083766\\
1.96	0.154697616879892\\
1.97	-0.14842864646848\\
1.98	-0.458345801379754\\
1.99	-0.763973205212376\\
2	-1.05439403672798\\
2.01	-1.31929002812936\\
2.02	-1.54935233747667\\
2.03	-1.73665218148987\\
2.04	-1.87495645297284\\
2.05	-1.95997574038237\\
2.06	-1.98953486197423\\
2.07	-1.96365911606249\\
2.08	-1.88457280905345\\
2.09	-1.75661012012623\\
2.1	-1.5860418562899\\
2.11	-1.38082500473222\\
2.12	-1.15028506720047\\
2.13	-0.904743840922086\\
2.14	-0.655107485444991\\
2.15	-0.41243129805069\\
2.16	-0.187478548946944\\
2.17	0.00970903573759918\\
2.18	0.170212019497947\\
2.19	0.286589252906719\\
2.2	0.353174930316012\\
2.21	0.366304862929974\\
2.22	0.32446294729257\\
2.23	0.22834198979626\\
2.24	0.0808164621723486\\
2.25	-0.11317172603453\\
2.26	-0.346809884270244\\
2.27	-0.611685787112853\\
2.28	-0.898122696169183\\
2.29	-1.19556480160181\\
2.3	-1.49299771708762\\
2.31	-1.77938726453547\\
2.32	-2.0441190557333\\
2.33	-2.27742134631986\\
2.34	-2.47075430435334\\
2.35	-2.61715017468899\\
2.36	-2.71149077800487\\
2.37	-2.75071128157986\\
2.38	-2.73392211823847\\
2.39	-2.66244419304784\\
2.4	-2.53975597428957\\
2.41	-2.37135457812737\\
2.42	-2.1645363851813\\
2.43	-1.92810593522931\\
2.44	-1.67202470557553\\
2.45	-1.40701377527091\\
2.46	-1.14412621579262\\
2.47	-0.894306255696073\\
2.48	-0.667952794029036\\
2.49	-0.474504663923524\\
2.5	-0.32206418066333\\
2.51	-0.217073982243682\\
2.52	-0.164060045829442\\
2.53	-0.165451125290965\\
2.54	-0.221481808325605\\
2.55	-0.330183058018966\\
2.56	-0.487460615970561\\
2.57	-0.687258141335595\\
2.58	-0.921798581982747\\
2.59	-1.1818941550925\\
2.6	-1.45731257927617\\
2.61	-1.73718495772032\\
2.62	-2.01043905136277\\
2.63	-2.26624066888137\\
2.64	-2.49442557668194\\
2.65	-2.68590471000287\\
2.66	-2.83302653065386\\
2.67	-2.92988208532669\\
2.68	-2.97254060276209\\
2.69	-2.95920623724725\\
2.7	-2.8902897095629\\
2.71	-2.76839198926424\\
2.72	-2.59820066880987\\
2.73	-2.38630316075068\\
2.74	-2.14092416518646\\
2.75	-1.87159787380286\\
2.76	-1.58878797863971\\
2.77	-1.30347063460169\\
2.78	-1.02669700163408\\
2.79	-0.769152806577233\\
2.8	-0.540732483524328\\
2.81	-0.350144870306913\\
2.82	-0.204566180681395\\
2.83	-0.109354087056676\\
2.84	-0.0678343123122336\\
2.85	-0.0811682385451528\\
2.86	-0.148306810689929\\
2.87	-0.266032572648333\\
2.88	-0.42908815999666\\
2.89	-0.630386126583268\\
2.9	-0.861291739801772\\
2.91	-1.11196747028712\\
2.92	-1.37176544222012\\
2.93	-1.62965219838189\\
2.94	-1.87464884580994\\
2.95	-2.09626903472767\\
2.96	-2.28493730979673\\
2.97	-2.43237115523106\\
2.98	-2.53191150272056\\
2.99	-2.57878852573126\\
3	-2.57031212367867\\
3.01	-2.50597950183899\\
3.02	-2.38749555799514\\
3.03	-2.21870526293043\\
3.04	-2.005440730407\\
3.05	-1.75528907332517\\
3.06	-1.47729030075957\\
3.07	-1.18157729961846\\
3.08	-0.878972253568907\\
3.09	-0.580555588579587\\
3.1	-0.297224629705457\\
3.11	-0.039259563917743\\
3.12	0.184085987490747\\
3.13	0.364953464500767\\
3.14	0.497192241374556\\
3.15	0.576604403981389\\
3.16	0.601111673117071\\
3.17	0.57083781927941\\
3.18	0.488103283461741\\
3.19	0.357332218591137\\
3.2	0.184875657756406\\
3.21	-0.0212421408683505\\
3.22	-0.251644059794467\\
3.23	-0.495974426526197\\
3.24	-0.743312333886917\\
3.25	-0.982607498398193\\
3.26	-1.20312128022476\\
3.27	-1.39485528241929\\
3.28	-1.54895044237498\\
3.29	-1.65804070477658\\
3.3	-1.71654717602781\\
3.31	-1.72090103295021\\
3.32	-1.66968629889091\\
3.33	-1.56369679500605\\
3.34	-1.40590499605188\\
3.35	-1.20134403210613\\
3.36	-0.956907540235491\\
3.37	-0.681075345190106\\
3.38	-0.38357590516752\\
3.39	-0.0749989796646459\\
3.4	0.233626039075372\\
3.41	0.531269320631158\\
3.42	0.807339807888881\\
3.43	1.05210693567468\\
3.44	1.25708804414784\\
3.45	1.41538604220892\\
3.46	1.52196385955533\\
3.47	1.57384474604985\\
3.48	1.57023043330439\\
3.49	1.51253244795214\\
3.5	1.40431532845281\\
3.51	1.25115400940078\\
3.52	1.06041105917491\\
3.53	0.840942651966823\\
3.54	0.602744996356472\\
3.55	0.356555316411236\\
3.56	0.113423293135235\\
3.57	-0.115729948255983\\
3.58	-0.320547723251312\\
3.59	-0.491651167596399\\
3.6	-0.621013631149331\\
3.61	-0.70228115994225\\
3.62	-0.731026289647293\\
3.63	-0.704925032310069\\
3.64	-0.623850000257805\\
3.65	-0.489875954435268\\
3.66	-0.307197555788519\\
3.67	-0.0819625985014588\\
3.68	0.177972626633452\\
3.69	0.463356086002186\\
3.7	0.763908405787787\\
3.71	1.06873223654916\\
3.72	1.36674625986218\\
3.73	1.64712653472258\\
3.74	1.89973758879343\\
3.75	2.11553606841994\\
3.76	2.2869308553274\\
3.77	2.4080852934486\\
3.78	2.47514947721488\\
3.79	2.48641334087176\\
3.8	2.44237444578309\\
3.81	2.3457177634186\\
3.82	2.20120826016126\\
3.83	2.01550056639061\\
3.84	1.79687331787984\\
3.85	1.55489876062212\\
3.86	1.30006079103816\\
3.87	1.04333665923479\\
3.88	0.795759011620063\\
3.89	0.567975732984754\\
3.9	0.36982513588647\\
3.91	0.209943433318817\\
3.92	0.095420143614264\\
3.93	0.0315151656177107\\
3.94	0.021448803565921\\
3.95	0.0662731128292079\\
3.96	0.164829695664451\\
3.97	0.313795629631627\\
3.98	0.50781569775616\\
3.99	0.739715648933815\\
4	1.00078798665397\\
};

\addplot [color=green!50!black,dashed, line width = 2pt]
  table[row sep=crcr]{%
0	0\\
0.01	1.38286659467745\\
0.02	1.8476294209268\\
0.03	1.9366722486999\\
0.04	1.8545698611844\\
0.05	1.68157828340957\\
0.06	1.45281562351316\\
0.07	1.18748889265735\\
0.08	0.899522093954995\\
0.09	0.601246752589555\\
0.1	0.304498522774308\\
0.11	0.0207448718846678\\
0.12	-0.239141210149401\\
0.13	-0.465222088819214\\
0.14	-0.648851771817584\\
0.15	-0.783001583900225\\
0.16	-0.862530385377584\\
0.17	-0.884384092954037\\
0.18	-0.847714807058356\\
0.19	-0.753914051772354\\
0.2	-0.606558415372018\\
0.21	-0.411269495963527\\
0.22	-0.175493514065382\\
0.23	0.0917908298598094\\
0.24	0.380424807802037\\
0.25	0.679468570382897\\
0.26	0.977630789580449\\
0.27	1.26371208334528\\
0.28	1.52704470332005\\
0.29	1.75791109902718\\
0.3	1.94792479565267\\
0.31	2.09035850571369\\
0.32	2.18040647887747\\
0.33	2.21537069647026\\
0.34	2.19476353448912\\
0.35	2.12032283131534\\
0.36	1.99593877213348\\
0.37	1.82749550293426\\
0.38	1.62263377356666\\
0.39	1.39044404672754\\
0.4	1.14110227296129\\
0.41	0.88546281058677\\
0.42	0.634624673016727\\
0.43	0.399488346194283\\
0.44	0.190320793431872\\
0.45	0.0163459386836471\\
0.46	-0.114623094950769\\
0.47	-0.196495376707652\\
0.48	-0.225144818546774\\
0.49	-0.19857444958129\\
0.5	-0.116995799479539\\
0.51	0.0171797449902901\\
0.52	0.199437241990534\\
0.53	0.423338297455182\\
0.54	0.680777775120769\\
0.55	0.962306870757294\\
0.56	1.25750966618861\\
0.57	1.55541815046831\\
0.58	1.84494916533235\\
0.59	2.11534586077264\\
0.6	2.35660606896899\\
0.61	2.5598805279979\\
0.62	2.71782508986057\\
0.63	2.82489288249398\\
0.64	2.87755478945838\\
0.65	2.8744394685394\\
0.66	2.81638733768299\\
0.67	2.70641638572873\\
0.68	2.54960117963835\\
0.69	2.35286989934495\\
0.7	2.12472749805662\\
0.71	1.87491602969414\\
0.72	1.61402568882158\\
0.73	1.35307207209588\\
0.74	1.10305651579725\\
0.75	0.874527037608603\\
0.76	0.677157385870573\\
0.77	0.519360976844485\\
0.78	0.407955109039064\\
0.79	0.347888838826445\\
0.8	0.342045363250055\\
0.81	0.391126785400132\\
0.82	0.493625853367693\\
0.83	0.645885796448748\\
0.84	0.842245870308887\\
0.85	1.07526680607479\\
0.86	1.33602717313113\\
0.87	1.61447883859395\\
0.88	1.89984735083514\\
0.89	2.18106128379461\\
0.9	2.44719342462525\\
0.91	2.6878962153968\\
0.92	2.89381408901784\\
0.93	3.05695626091044\\
0.94	3.17101511482182\\
0.95	3.23161749025584\\
0.96	3.23649885424847\\
0.97	3.18559341464153\\
0.98	3.08103658324228\\
0.99	2.92707969162697\\
1	2.72992036053945\\
1.01	2.49745528643301\\
1.02	2.23896530162616\\
1.03	1.96474526447691\\
1.04	1.68569353535794\\
1.05	1.41287740526458\\
1.06	1.15709180247393\\
1.07	0.928428870534927\\
1.08	0.735875577351233\\
1.09	0.586955397490975\\
1.1	0.487428352693363\\
1.11	0.441061368923738\\
1.12	0.44947810491505\\
1.13	0.512094238821711\\
1.14	0.626140792627226\\
1.15	0.786774564082586\\
1.16	0.987271263290682\\
1.17	1.2192936538342\\
1.18	1.47322400815365\\
1.19	1.73854762290536\\
1.2	2.00427210444672\\
1.21	2.25936570856457\\
1.22	2.49319725898352\\
1.23	2.69596010627186\\
1.24	2.85906322504861\\
1.25	2.97547385754022\\
1.26	3.03999804324701\\
1.27	3.04948785078955\\
1.28	3.0029670502027\\
1.29	2.90167021547512\\
1.3	2.74899369839491\\
1.31	2.55036042824428\\
1.32	2.31300392730491\\
1.33	2.04568015281373\\
1.34	1.75831865331816\\
1.35	1.46162694670802\\
1.36	1.16666389211537\\
1.37	0.884399063961172\\
1.38	0.625275694482641\\
1.39	0.39879460878906\\
1.4	0.213135739561829\\
1.41	0.0748323103482104\\
1.42	-0.0114893233954364\\
1.43	-0.0432938038884917\\
1.44	-0.0202375307134072\\
1.45	0.055817966218621\\
1.46	0.180880573413147\\
1.47	0.348987088134785\\
1.48	0.55244130163399\\
1.49	0.782121179457591\\
1.5	1.02784289233936\\
1.51	1.27876718472107\\
1.52	1.52383188007746\\
1.53	1.75219328022938\\
1.54	1.95365886125689\\
1.55	2.11909401558467\\
1.56	2.24078662454749\\
1.57	2.31275492685772\\
1.58	2.33098640905811\\
1.59	2.29359819397078\\
1.6	2.20091253276011\\
1.61	2.05544439080172\\
1.62	1.86180162209566\\
1.63	1.62650171175944\\
1.64	1.35771239235036\\
1.65	1.06492647460047\\
1.66	0.758583855899226\\
1.67	0.449655775653075\\
1.68	0.14920789177507\\
1.69	-0.132040403122011\\
1.7	-0.384142857333648\\
1.71	-0.598322409476768\\
1.72	-0.767320622657099\\
1.73	-0.885686572740983\\
1.74	-0.949993673863893\\
1.75	-0.958975797217128\\
1.76	-0.913577256686186\\
1.77	-0.816914668773033\\
1.78	-0.674152207523775\\
1.79	-0.492295227861605\\
1.8	-0.2799104851167\\
1.81	-0.0467841049268025\\
1.82	0.196469060630598\\
1.83	0.438829116717001\\
1.84	0.669310958083137\\
1.85	0.877402715156259\\
1.86	1.05348533418734\\
1.87	1.1892165639092\\
1.88	1.27786404017139\\
1.89	1.31457418928007\\
1.9	1.29656622954025\\
1.91	1.22324353652557\\
1.92	1.09621793212352\\
1.93	0.91924592878109\\
1.94	0.698079470521981\\
1.95	0.440237120975587\\
1.96	0.154704820265622\\
1.97	-0.14842185959955\\
1.98	-0.458339698875743\\
1.99	-0.763968028127112\\
2	-1.05438998973975\\
2.01	-1.31928727138867\\
2.02	-1.5493509802148\\
2.03	-1.73665227763781\\
2.04	-1.87495799897356\\
2.05	-1.95997867528068\\
2.06	-1.98953906978963\\
2.07	-1.96366443034826\\
2.08	-1.88457901946322\\
2.09	-1.75661698073134\\
2.1	-1.58604909531662\\
2.11	-1.38083233533114\\
2.12	-1.15029219882583\\
2.13	-0.904750490862455\\
2.14	-0.655113390052625\\
2.15	-0.412436223213414\\
2.16	-0.187482299400554\\
2.17	0.00970660864416585\\
2.18	0.170211011882221\\
2.19	0.286589704520684\\
2.2	0.353176822951296\\
2.21	0.366308121129821\\
2.22	0.324467441339342\\
2.23	0.22834754085395\\
2.24	0.080822849390924\\
2.25	-0.113164756742958\\
2.26	-0.346802610138675\\
2.27	-0.611678497496177\\
2.28	-0.89811568103409\\
2.29	-1.19555833999798\\
2.3	-1.49299206604657\\
2.31	-1.77938264884395\\
2.32	-2.04411565898754\\
2.33	-2.27741930362125\\
2.34	-2.47075369693043\\
2.35	-2.61715102666235\\
2.36	-2.71149305542447\\
2.37	-2.75071489378125\\
2.38	-2.73392692144486\\
2.39	-2.66244999609596\\
2.4	-2.53976254623554\\
2.41	-2.3713616574436\\
2.42	-2.16454369017389\\
2.43	-1.9281131752392\\
2.44	-1.67203159256366\\
2.45	-1.4070200352815\\
2.46	-1.14413159986177\\
2.47	-0.894310549765463\\
2.48	-0.667955827466265\\
2.49	-0.474506316317156\\
2.5	-0.322064386614042\\
2.51	-0.217072733966537\\
2.52	-0.164057393458361\\
2.53	-0.165447174878552\\
2.54	-0.221476717612555\\
2.55	-0.330177030148691\\
2.56	-0.487453891388807\\
2.57	-0.687250988213068\\
2.58	-0.921791285520945\\
2.59	-1.18188700616262\\
2.6	-1.45730586282499\\
2.61	-1.73717894141692\\
2.62	-2.01043397493318\\
2.63	-2.26623673454949\\
2.64	-2.49442294112222\\
2.65	-2.68590347809035\\
2.66	-2.8330267512884\\
2.67	-2.92988374948606\\
2.68	-2.97254364386192\\
2.69	-2.95921053379881\\
2.7	-2.89029509001326\\
2.71	-2.76839823884072\\
2.72	-2.59820753807616\\
2.73	-2.38631037555462\\
2.74	-2.140931437583\\
2.75	-1.8716049135373\\
2.76	-1.58879450470785\\
2.77	-1.30347638645849\\
2.78	-1.02670174957801\\
2.79	-0.769156360897671\\
2.8	-0.540734702073895\\
2.81	-0.350145664157408\\
2.82	-0.204565517670847\\
2.83	-0.10935199307174\\
2.84	-0.0678308702508112\\
2.85	-0.0811635850167003\\
2.86	-0.148301130564295\\
2.87	-0.266026091686937\\
2.88	-0.429081135855812\\
2.89	-0.630378838538334\\
2.9	-0.861284477617576\\
2.91	-1.11196052266777\\
2.92	-1.37175908529598\\
2.93	-1.6296466847105\\
2.94	-1.87464439430208\\
2.95	-2.09626582192809\\
2.96	-2.28493546284173\\
2.97	-2.4323707467887\\
2.98	-2.53191254808962\\
2.99	-2.57879098223443\\
3	-2.57031589237253\\
3.01	-2.5059844314505\\
3.02	-2.38750145096041\\
3.03	-2.2187118832692\\
3.04	-2.00544781313386\\
3.05	-1.75529633501824\\
3.06	-1.47729745085492\\
3.07	-1.1815840519955\\
3.08	-0.878978337958513\\
3.09	-0.580560761349051\\
3.1	-0.297228683559691\\
3.11	-0.0392623361697111\\
3.12	0.184084608434093\\
3.13	0.36495353468604\\
3.14	0.497193759068597\\
3.15	0.576607309739082\\
3.16	0.601115852151258\\
3.17	0.57084310603507\\
3.18	0.48810946821623\\
3.19	0.357339055811146\\
3.2	0.184882875887072\\
3.21	-0.0212348285797701\\
3.22	-0.251636943861379\\
3.23	-0.495967789653586\\
3.24	-0.743306439692799\\
3.25	-0.98260258091127\\
3.26	-1.20311753454796\\
3.27	-1.39485285696247\\
3.28	-1.54894943293363\\
3.29	-1.65804115071682\\
3.3	-1.71654905871937\\
3.31	-1.7209042765063\\
3.32	-1.66969077319448\\
3.33	-1.56370232090417\\
3.34	-1.40591135249482\\
3.35	-1.20135096496346\\
3.36	-0.956914772424001\\
3.37	-0.681082587728152\\
3.38	-0.383582868693724\\
3.39	-0.0750053859735767\\
3.4	0.233620445943041\\
3.41	0.531264764180526\\
3.42	0.807336470258263\\
3.43	1.05210495037698\\
3.44	1.25708749074337\\
3.45	1.41538694313843\\
3.46	1.5219661792398\\
3.47	1.57384839231066\\
3.48	1.57023526103779\\
3.49	1.51253826491202\\
3.5	1.40432190291657\\
3.51	1.25116107940585\\
3.52	1.06041834296054\\
3.53	0.84094985921162\\
3.54	0.602751839745921\\
3.55	0.356561523096777\\
3.56	0.113428615609464\\
3.57	-0.11572572229248\\
3.58	-0.32054476242457\\
3.59	-0.491649590138235\\
3.6	-0.621013500184494\\
3.61	-0.702282480969099\\
3.62	-0.731029010323232\\
3.63	-0.704929044529638\\
3.64	-0.623855144473545\\
3.65	-0.489882026012248\\
3.66	-0.307204313158818\\
3.67	-0.0819697728031615\\
3.68	0.177965320844644\\
3.69	0.463348939372527\\
3.7	0.763901702575049\\
3.71	1.06872624329344\\
3.72	1.36674121476065\\
3.73	1.64712263813334\\
3.74	1.89973499524601\\
3.75	2.11553488045836\\
3.76	2.28693111941975\\
3.77	2.40808699813803\\
3.78	2.4751525535765\\
3.79	2.48641766525901\\
3.8	2.44237984476128\\
3.81	2.34572402067443\\
3.82	2.20121512513345\\
3.83	2.01550776425732\\
3.84	1.79688056051195\\
3.85	1.55490575807824\\
3.86	1.30006726312074\\
3.87	1.04334234665977\\
3.88	0.795763686357798\\
3.89	0.567979207350829\\
3.9	0.369827270026545\\
3.91	0.209944140779308\\
3.92	0.0954193947976906\\
3.93	0.0315129889583883\\
3.94	0.021445284400294\\
3.95	0.0662683899931081\\
3.96	0.16482395596105\\
3.97	0.313789100385329\\
3.98	0.507808637745689\\
3.99	0.739708338085648\\
4	1.00078071487798\\
};

\addplot [color=magenta,dashdotted, line width = 2pt]
  table[row sep=crcr]{%
0	0\\
0.01	1.38290999797577\\
0.02	1.84755856371577\\
0.03	1.93646439890139\\
0.04	1.85433440221251\\
0.05	1.68142568674508\\
0.06	1.45279819933827\\
0.07	1.18759986395809\\
0.08	0.899718335294433\\
0.09	0.601473855605112\\
0.1	0.304709269519978\\
0.11	0.0209091906045572\\
0.12	-0.239033619916283\\
0.13	-0.465164262694826\\
0.14	-0.64882489197563\\
0.15	-0.782981440392491\\
0.16	-0.862493521486565\\
0.17	-0.884312747803828\\
0.18	-0.847600184510466\\
0.19	-0.753757778392319\\
0.2	-0.60637229961046\\
0.21	-0.411073871090043\\
0.22	-0.175314548042439\\
0.23	0.0919244325173884\\
0.24	0.380485292505413\\
0.25	0.679432428772108\\
0.26	0.97748155325883\\
0.27	1.26344238387555\\
0.28	1.52665745497358\\
0.29	1.7574197475812\\
0.3	1.94735265662664\\
0.31	2.08973728440454\\
0.32	2.17977411939806\\
0.33	2.21476874600229\\
0.34	2.1942342316029\\
0.35	2.11990613397256\\
0.36	1.99566953316996\\
0.37	1.82740098040135\\
0.38	1.62273163271331\\
0.39	1.39074097232044\\
0.4	1.14159326791752\\
0.41	0.886131212287061\\
0.42	0.635442874812252\\
0.43	0.400419170849587\\
0.44	0.191319429539424\\
0.45	0.0173623220124018\\
0.46	-0.113641594547524\\
0.47	-0.195601115137128\\
0.48	-0.224387040756643\\
0.49	-0.197996608967912\\
0.5	-0.116633174414667\\
0.51	0.0173020150122813\\
0.52	0.199305602124624\\
0.53	0.422951634940159\\
0.54	0.680147648950866\\
0.55	0.961457106051702\\
0.56	1.2564753371245\\
0.57	1.55424400510381\\
0.58	1.84368757186365\\
0.59	2.11405437662757\\
0.6	2.35534475078461\\
0.61	2.55870911133907\\
0.62	2.71680017217683\\
0.63	2.82406524124926\\
0.64	2.87696695981414\\
0.65	2.87412369189616\\
0.66	2.81636397450217\\
0.67	2.70669286423676\\
0.68	2.55017152731574\\
0.69	2.35371487762608\\
0.7	2.12581533362211\\
0.71	1.87620370929656\\
0.72	1.61546075985932\\
0.73	1.35459486922527\\
0.74	1.10460271557446\\
0.75	0.876030429309523\\
0.76	0.678552737675887\\
0.77	0.520586872913913\\
0.78	0.408956634678117\\
0.79	0.348619997836292\\
0.8	0.342471123356837\\
0.81	0.391224663794963\\
0.82	0.49338697413581\\
0.83	0.645315374228299\\
0.84	0.841363098854851\\
0.85	1.07410415555752\\
0.86	1.33462912485469\\
0.87	1.61290010942109\\
0.88	1.89815068090403\\
0.89	2.17931487931074\\
0.9	2.44546816182747\\
0.91	2.68626272168911\\
0.92	2.89233982229188\\
0.93	3.05570270817592\\
0.94	3.17003522626533\\
0.95	3.23095345517873\\
0.96	3.23618031122642\\
0.97	3.18563617041553\\
0.98	3.08144189397517\\
0.99	2.92783413710841\\
1	2.73099631765927\\
1.01	2.49881198373437\\
1.02	2.24055041300279\\
1.03	1.96649697807428\\
1.04	1.6875430143229\\
1.05	1.41475154099817\\
1.06	1.15891614908134\\
1.07	0.930130641729804\\
1.08	0.737386584447725\\
1.09	0.588214809417061\\
1.1	0.488385166081825\\
1.11	0.441676487960265\\
1.12	0.449725946338455\\
1.13	0.511963796577582\\
1.14	0.625636118419654\\
1.15	0.785914643621471\\
1.16	0.986089292342306\\
1.17	1.21783574236854\\
1.18	1.47154736393259\\
1.19	1.73671828682808\\
1.2	2.00236232804106\\
1.21	2.25745107847147\\
1.22	2.49135368350861\\
1.23	2.69426078469746\\
1.24	2.85757572118231\\
1.25	2.97425739486676\\
1.26	3.03910113024369\\
1.27	3.04894633180985\\
1.28	3.00280266032824\\
1.29	2.9018896975991\\
1.3	2.74958851834042\\
1.31	2.55130709976314\\
1.32	2.31426493444469\\
1.33	2.04720543321228\\
1.34	1.76004758352885\\
1.35	1.46349075056386\\
1.36	1.1685883761807\\
1.37	0.886307571011478\\
1.38	0.62709215718364\\
1.39	0.400446581693174\\
1.4	0.214557288367386\\
1.41	0.0759666430012637\\
1.42	-0.0106875888825663\\
1.43	-0.0428568255798658\\
1.44	-0.0201829554272984\\
1.45	0.0554877121234613\\
1.46	0.180178386624053\\
1.47	0.347940680036229\\
1.48	0.55109209909875\\
1.49	0.780522678431856\\
1.5	1.02605852972938\\
1.51	1.27686781332329\\
1.52	1.5218929471535\\
1.53	1.75029182217282\\
1.54	1.95187043415526\\
1.55	2.11748968407416\\
1.56	2.23943012903298\\
1.57	2.31170014210677\\
1.58	2.3302751955596\\
1.59	2.29325872779378\\
1.6	2.20095818059504\\
1.61	2.05587317512806\\
1.62	1.8625962978848\\
1.63	1.62763045175371\\
1.64	1.35913005365977\\
1.65	1.0665763962212\\
1.66	0.760400115540695\\
1.67	0.451565815957686\\
1.68	0.15113541127032\\
1.69	-0.130172409558854\\
1.7	-0.382409029696403\\
1.71	-0.596792047871836\\
1.72	-0.766054925161242\\
1.73	-0.884736196004564\\
1.74	-0.94939671372589\\
1.75	-0.958756269864881\\
1.76	-0.913744140937831\\
1.77	-0.817461547692712\\
1.78	-0.675057523692068\\
1.79	-0.493523142154803\\
1.8	-0.28141230479041\\
1.81	-0.0485002240966716\\
1.82	0.19460678543794\\
1.83	0.43689465062906\\
1.84	0.667381139819586\\
1.85	0.87555419435049\\
1.86	1.05179151618026\\
1.87	1.18774468381042\\
1.88	1.27667248282107\\
1.89	1.31371016197177\\
1.9	1.29606388026996\\
1.91	1.2231225928097\\
1.92	1.09648291463719\\
1.93	0.919885971195505\\
1.94	0.699068752713985\\
1.95	0.441535898370541\\
1.96	0.156261008266318\\
1.97	-0.146670609240959\\
1.98	-0.456463512512053\\
1.99	-0.762042014584384\\
2	-1.05249124603821\\
2.01	-1.31749180918591\\
2.02	-1.54773069549302\\
2.03	-1.73527208444676\\
2.04	-1.87387324148866\\
2.05	-1.95923292135296\\
2.06	-1.98916237394506\\
2.07	-1.96367213549025\\
2.08	-1.88497114505221\\
2.09	-1.7573782218322\\
2.1	-1.58714943255722\\
2.11	-1.38222823145925\\
2.12	-1.15192833416052\\
2.13	-0.906561968463128\\
2.14	-0.6570283226906\\
2.15	-0.414378599024921\\
2.16	-0.189375011964983\\
2.17	0.00793868658041524\\
2.18	0.168638033585373\\
2.19	0.285274052737067\\
2.2	0.352170623154428\\
2.21	0.365651163679405\\
2.22	0.3241855953918\\
2.23	0.228451723259027\\
2.24	0.0813085896703204\\
2.25	-0.112317137984254\\
2.26	-0.345627216473616\\
2.27	-0.610222496909326\\
2.28	-0.896437425331136\\
2.29	-1.19372503835343\\
2.3	-1.49107710544888\\
2.31	-1.77746266825808\\
2.32	-2.04226749387829\\
2.33	-2.27571692264435\\
2.34	-2.46926525292154\\
2.35	-2.61593613948816\\
2.36	-2.71060043501936\\
2.37	-2.750180398123\\
2.38	-2.73377212691079\\
2.39	-2.66268133717342\\
2.4	-2.54037105889971\\
2.41	-2.37232333641968\\
2.42	-2.16582044584499\\
2.43	-1.92965435207371\\
2.44	-1.67377598848065\\
2.45	-1.4088983414936\\
2.46	-1.14606916390326\\
2.47	-0.896230351558472\\
2.48	-0.669781549782398\\
2.49	-0.47616538721831\\
2.5	-0.323490872587797\\
2.51	-0.218209968410606\\
2.52	-0.164860235729854\\
2.53	-0.165883809842585\\
2.54	-0.22152992395265\\
2.55	-0.329844866868977\\
2.56	-0.486749775150247\\
2.57	-0.686203158404968\\
2.58	-0.920441678415192\\
2.59	-1.18028958301098\\
2.6	-1.45552445857148\\
2.61	-1.73528471977825\\
2.62	-2.0085025913108\\
2.63	-2.26434531988493\\
2.64	-2.4926470269157\\
2.65	-2.68431398520682\\
2.66	-2.83168716257056\\
2.67	-2.92884757890413\\
2.68	-2.971852303097\\
2.69	-2.95889168130712\\
2.7	-2.89036152839744\\
2.71	-2.76884740447261\\
2.72	-2.5990216033367\\
2.73	-2.38745695959474\\
2.74	-2.14236489728886\\
2.75	-1.87326816321314\\
2.76	-1.59062129195656\\
2.77	-1.3053939334957\\
2.78	-1.02863365470458\\
2.79	-0.771025644456611\\
2.8	-0.542466875433135\\
2.81	-0.351671699406174\\
2.82	-0.2058245996055\\
2.83	-0.110293943771181\\
2.84	-0.0684181495862458\\
2.85	-0.0813727873252509\\
2.86	-0.148123917839634\\
2.87	-0.265469526066375\\
2.88	-0.428167398153521\\
2.89	-0.629144344055195\\
2.9	-0.859778424461691\\
2.91	-1.11024293047024\\
2.92	-1.36989840251428\\
2.93	-1.62771705989144\\
2.94	-1.87272272011698\\
2.95	-2.09442866979748\\
2.96	-2.28325603038195\\
2.97	-2.43091593975018\\
2.98	-2.53074031314704\\
2.99	-2.57794799694128\\
3	-2.56983570437649\\
3.01	-2.50588612117484\\
3.02	-2.38778887103312\\
3.03	-2.2193735050475\\
3.04	-2.00645718642764\\
3.05	-1.75661314268828\\
3.06	-1.47886911625278\\
3.07	-1.18334783514635\\
3.08	-0.880863836924552\\
3.09	-0.582492719043364\\
3.1	-0.299129988141778\\
3.11	-0.0410570953674914\\
3.12	0.182468041616472\\
3.13	0.363579705508782\\
3.14	0.496117537715601\\
3.15	0.575871703694185\\
3.16	0.600750291500972\\
3.17	0.570862270061284\\
3.18	0.488512700054562\\
3.19	0.358110388517378\\
3.2	0.185991668863058\\
3.21	-0.0198326681647384\\
3.22	-0.249997203358342\\
3.23	-0.494155726980618\\
3.24	-0.741394181839909\\
3.25	-0.980666248581813\\
3.26	-1.2012342075965\\
3.27	-1.39309750157666\\
3.28	-1.54739191309609\\
3.29	-1.65674344306664\\
3.3	-1.71556278185778\\
3.31	-1.72026863324947\\
3.32	-1.66943098779977\\
3.33	-1.5638286335701\\
3.34	-1.40641861123774\\
3.35	-1.20221883112392\\
3.36	-0.958108531623989\\
3.37	-0.682554533930399\\
3.38	-0.385274206206032\\
3.39	-0.0768485735595947\\
3.4	0.231699002296286\\
3.41	0.529341777196128\\
3.42	0.805488712969715\\
3.43	1.05040619532068\\
3.44	1.25560556878022\\
3.45	1.41418103914335\\
3.46	1.52108447249177\\
3.47	1.57332613563517\\
3.48	1.5700933752901\\
3.49	1.51278250483448\\
3.5	1.4049426276188\\
3.51	1.25213363662532\\
3.52	1.06170405178121\\
3.53	0.842497552056452\\
3.54	0.604499902185963\\
3.55	0.358440350133724\\
3.56	0.115363386526824\\
3.57	-0.113812061144616\\
3.58	-0.318728425824707\\
3.59	-0.490002915626122\\
3.6	-0.619602064266481\\
3.61	-0.701162484865341\\
3.62	-0.730245039468849\\
3.63	-0.704512291171633\\
3.64	-0.62382216418981\\
3.65	-0.490234077763604\\
3.66	-0.307927309164207\\
3.67	-0.08303484022656\\
3.68	0.176600688760477\\
3.69	0.461739188640919\\
3.7	0.762111047824544\\
3.71	1.06682610767905\\
3.72	1.36480738249837\\
3.73	1.64523223317609\\
3.74	1.89796340655279\\
3.75	2.11395275642883\\
3.76	2.28560155132818\\
3.77	2.40706300484258\\
3.78	2.47447496778949\\
3.79	2.48611350564449\\
3.8	2.44246123875157\\
3.81	2.34618772096452\\
3.82	2.20204263910814\\
3.83	2.0166660912117\\
3.84	1.79832350726722\\
3.85	1.55657578053113\\
3.86	1.30189776033141\\
3.87	1.04526031600802\\
3.88	0.79769263391083\\
3.89	0.569842197425471\\
3.9	0.371549992361859\\
3.91	0.211457873051955\\
3.92	0.0966637423523043\\
3.93	0.0324382925551328\\
3.94	0.0220145999694693\\
3.95	0.0664589614827782\\
3.96	0.164628122551659\\
3.97	0.313214601908995\\
3.98	0.506878306137559\\
3.99	0.73845918714854\\
4	0.999262464677408\\
};

\end{axis}
\end{tikzpicture}%